\documentclass[preprint2]{aastex63}

\usepackage{amsmath}
\usepackage{CJK}

\shorttitle{CMR exploration I}

\begin{document}
\begin{CJK*}{UTF8}{gbsn}

\title{CMR exploration I -- filament structure with synthetic observations}

\author[0000-0002-8469-2029]{Shuo Kong (孔朔)}
\affiliation{Steward Observatory, University of Arizona, Tucson, AZ 85719, USA}

\author[0000-0002-8351-3877]{Volker Ossenkopf-Okada}
\affiliation{I.~Physikalisches Institut, Universit\"at zu K\"oln,
              Z\"ulpicher Str. 77, 50937 K\"oln, Germany}

\author[0000-0001-5653-7817]{H\'ector G. Arce}
\affiliation{Department of Astronomy, Yale University, New Haven, CT 06511, USA}

\author[0000-0002-0560-3172]{Ralf S. Klessen}
\affiliation{Universit\"{a}t Heidelberg, Zentrum f\"{u}r Astronomie, Institut f\"{u}r Theoretische Astrophysik, Albert-Ueberle-Str. 2, 69120 Heidelberg, Germany}
\affiliation{Universit\"{a}t Heidelberg, Interdisziplin\"{a}res Zentrum f\"{u}r Wissenschaftliches Rechnen, Im Neuenheimer Feld 205, 69120 Heidelberg, Germany}

\author[0000-0001-6216-8931]{Duo Xu}
\affiliation{Department of Astronomy, University of Virginia, Charlottesville, VA 22904, USA}

\begin{abstract}
In this paper, we carry out a pilot parameter exploration for the collision-induced magnetic reconnection (CMR) mechanism that forms filamentary molecular clouds. Following \citet{2021ApJ...906...80K}, we utilize Athena++ to model CMR in the context of resistive magnetohydrodynamics (MHD), considering the effect from seven physical conditions, including the Ohmic resistivity ($\eta$), the magnetic field ($B$), the cloud density ($\rho$), the cloud radius $R$, the isothermal temperature $T$, the collision velocity $v_x$, and the shear velocity $v_z$. Compared to their fiducial model, we consider a higher and a lower value for each one of the seven parameters. We quantify the exploration results with five metrics, including the density probability distribution function ($\rho$-PDF), the filament morphology (250 $\mu$m dust emission), the $B$-$\rho$ relation, the dominant fiber width, and the ringiness that describes the significance of the ring-like sub-structures. The exploration forms straight and curved CMR-filaments with rich sub-structures that are highly variable in space and time. The variation translates to fluctuation in all the five metrics, reflecting the chaotic nature of magnetic reconnection in CMR. A temporary $B\propto\rho$ relation is noticeable during the first 0.6 Myr. Overall, the exploration provides useful initial insights to the CMR mechanism.
\end{abstract}

\keywords{}

\section{Introduction}\label{sec:intro}
\end{CJK*}

For 25 years, we are aware that the Orion A giant molecular cloud sits between a reverse magnetic field \citep[][hereafter H97]{1997ApJS..111..245H}. However, it is not clear if the field-reversal means anything. Was it dynamically involved in the formation of the cloud? Or was it simply a coincidence without any meaningful role? As shown in H97 (see their Figure 15), the field-reversal is quite remarkable \citep[also see][]{2019A&A...632A..68T}. In Galactic coordinates, the field points toward us for $b\ga-19$ deg and away from us for $b\la-20$ deg. Between $b\sim-19$ and $b\sim-20$, the line-of-sight field direction is not clear, yet this location is where Orion A resides. Such a large-scale field-reversal and its relation to the giant cloud are hard to reconcile in previous physical models for the formation of the Orion A cloud.

Recently, when explaining a special Stick filament in the Orion A, \citet[][hereafter K21]{2021ApJ...906...80K} found that a collision-induced magnetic reconnection (CMR) mechanism was able to link the filament formation to the field-reversal and put these puzzle pieces into a single picture. Simply speaking, if two clouds collide at a reverse B-field, the filament formation and its follow-up features are automatically satisfied (see below). Based on the modeling for the Stick filament, K21 speculated that the entire Orion A cloud formed in a similar way. 

\begin{figure*}[htb!]
\centering
\includegraphics[width=0.49\textwidth]{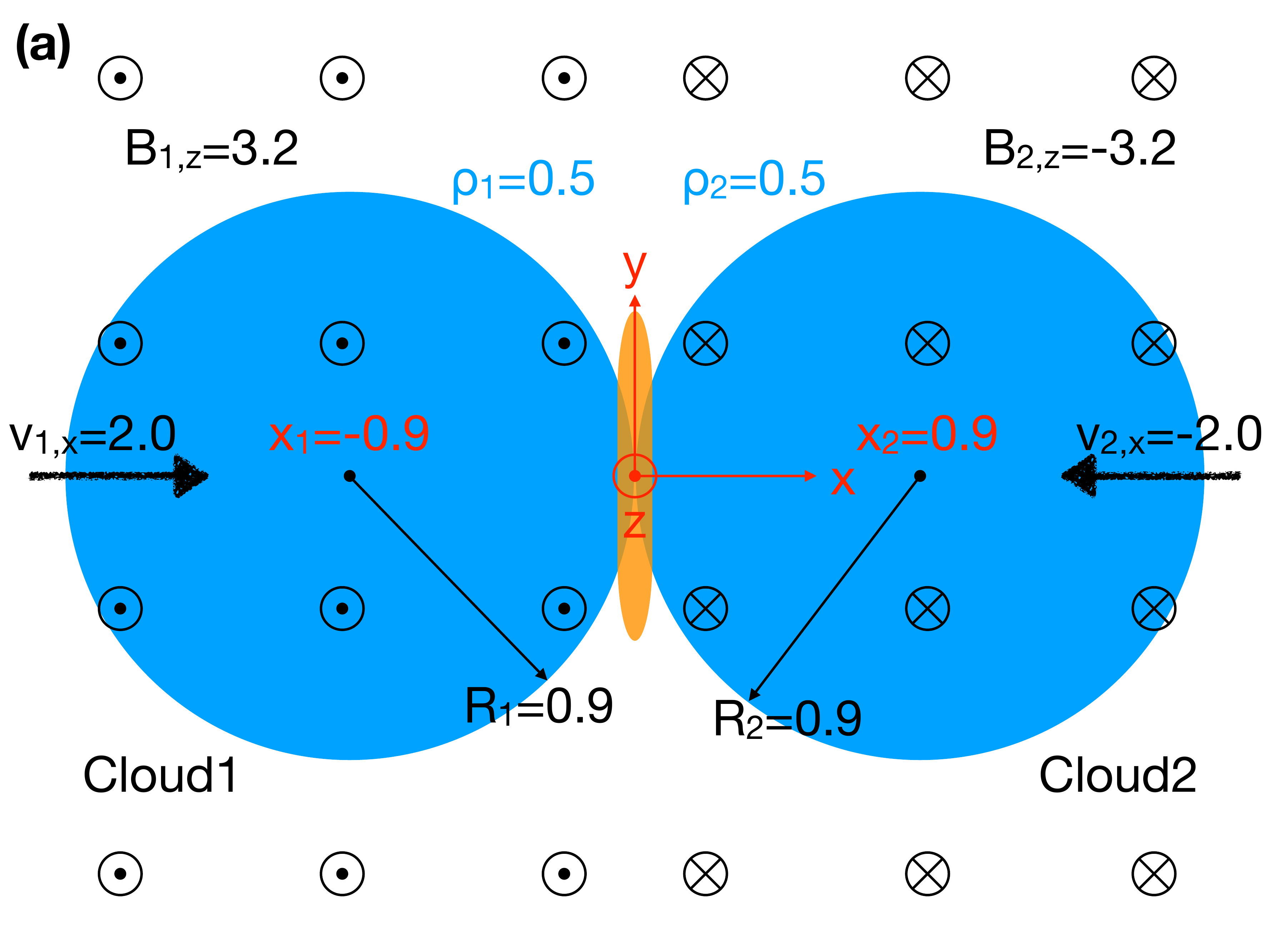}
\includegraphics[width=0.49\textwidth]{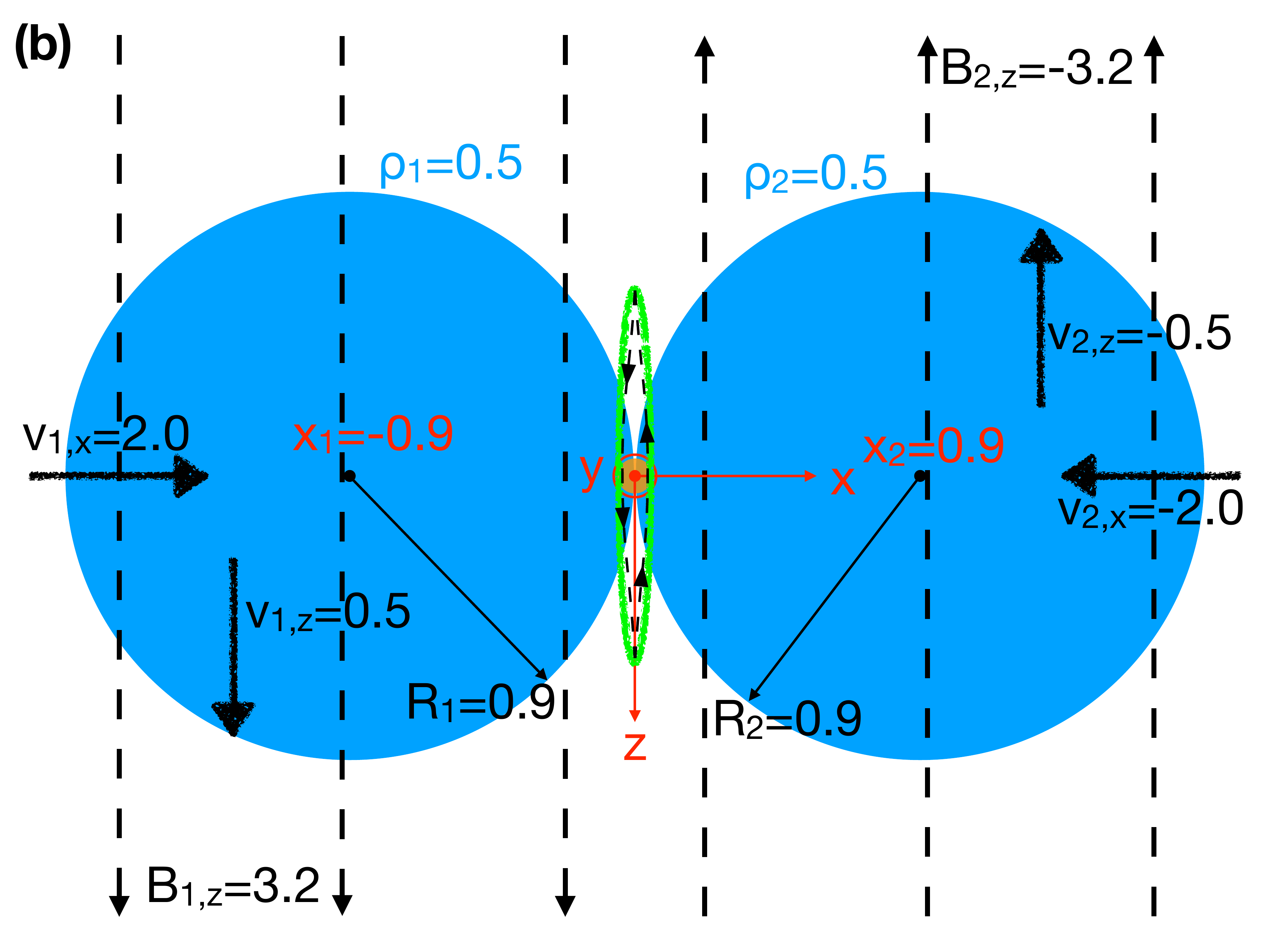}
\caption{
An illustration of CMR in two viewing angles. {\bf (a):} A view in the x-y plane. The Cartesian coordinate system (red) centers at the collision point. The x-axis points rightward and the y-axis points to the top. The z-axis points toward us as indicated by the red circle-point. The clouds have colliding velocities $v_{\rm 1,x}$ and $v_{\rm 2,x}$, respectively. The B-field points toward us (marked as black circle-points) for $x<0$ and away from us (marked as black circle-crosses) for $x>0$. After collision, the filament (orange) forms along the y-axis. {\bf (b):} A view in the z-x projection. In this view, the B-field is parallel to the plane of the sky. The y-axis points toward us as indicated by the red circle-point. After collision, the filament (orange) forms along the y-axis which points toward us. The green ellipse marks the location of the compression pancake if no B-fields. With antiparallel fields and CMR, the field reconnects at two tips of the pancake and forms a loop (black dashed arrow curve) around the pancake. Due to the magnetic tension force, the pancake is squeezed into the central axis (y-axis) becoming a filament. The fiducial model parameters in code unit are shown in both panels (see row 1 in Table \ref{tab:ic}).}
\label{fig:cmr}
\end{figure*}

Figure \ref{fig:cmr} illustrates the CMR filament formation. In panel (a), we view the process from the side of the filament. Two clouds move along the x-axis and collide at the origin. On the left side of the y-z plane, the B-field points toward us. On the other side, the field points away from us. After collision, the reverse field reconnects in the z-x plane and forms field loops that pull the compression pancake into the central axis (y-axis in our setup). The pulling is due to the magnetic tension the field loop exerts on the gas. As a result, a filamentary structure forms along the y-axis. In panel (b), we view the process in the z-x plane. In this projection, we are looking at the filament cross-section at the origin. The green ellipse represents the compression pancake and the black dashed arrow curve around the pancake denotes the reconnected field loop. The loop has a strong magnetic tension that pulls the dense gas in the pancake to the origin in each z-x plane. As a result, the filament (orange cross-section) forms along the y-axis. Essentially, the filament forms along the field symmetry axis that crosses the collision point. 

\begin{figure*}[htb!]
\centering
\epsscale{1.15}
\plotone{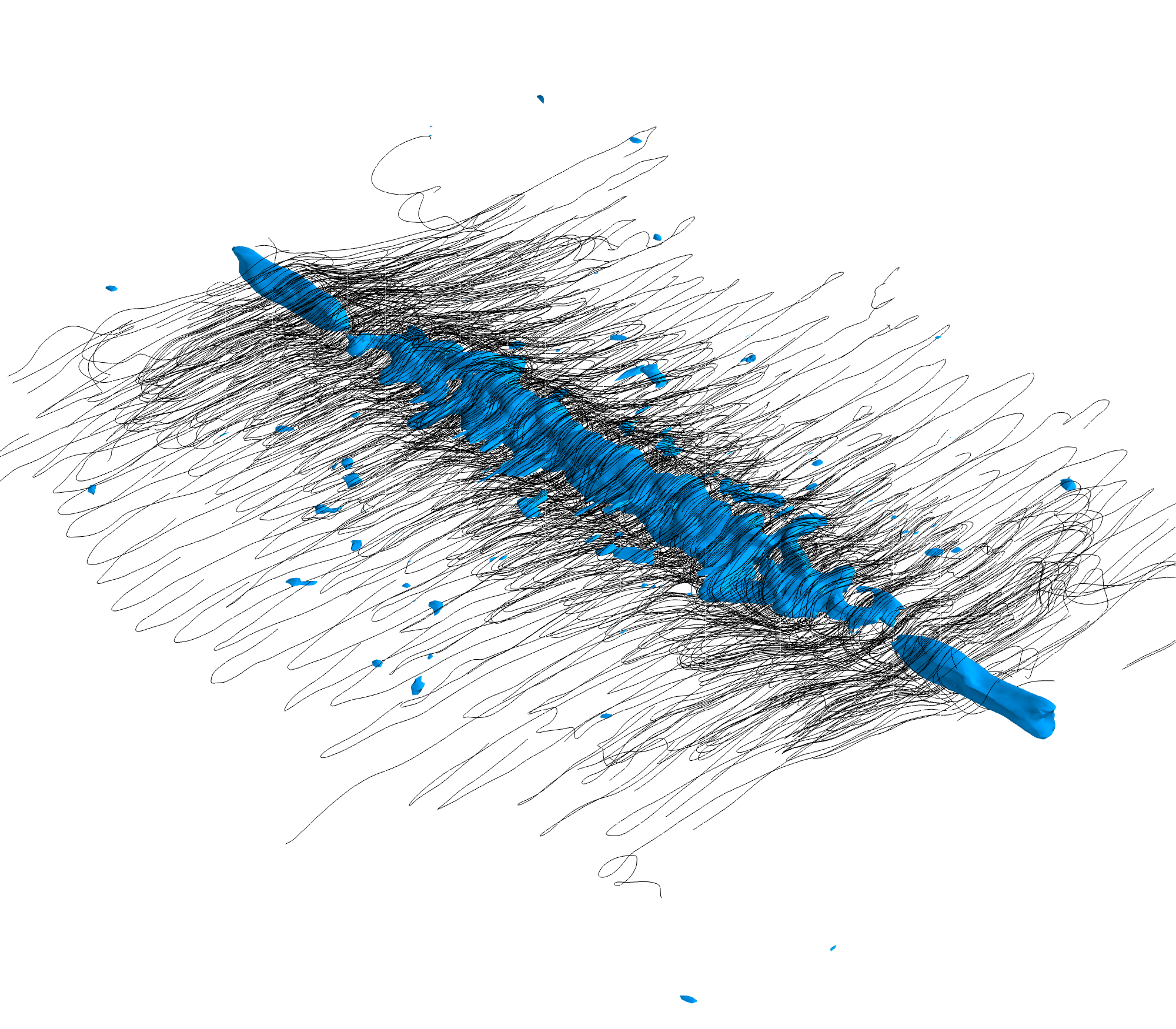}
\caption{
3D illustration of the field topology around the CMR-filament. The color plot shows the isosurface of the filament at a density of 10 ($n_{\rm H_2}=8400$ cm$^{-3}$). The black lines are B-field streamlines. The streamline seeding and length are limited so as to avoid too much crowdedness. See Figure \ref{fig:mrcolrho} for an illustration of the 3D geometry.
\label{fig:stream}}
\end{figure*}

A natural result of the CMR process is a helical B-field wrapping around the filament, exerting a surface magnetic pressure on the filament. Figure \ref{fig:stream} shows the field topology around the filament at t=0.6 Myr in the fiducial model \mbox{MRCOL} in K21. One can see multiple field loops closing in on the filament. With the progress of the collision, field loops keep forming at the two ends of the collision pancake, and they continue pulling material to the filament, which is why we see loops at different radii from the filament. Once the loops become circular around the filament, they stop pulling the gas and become a confining force over the filament surface. Such confinement helps preserving the dense gas until gravity takes over. In the CMR scenario, B-field is proactive in creating dense gas rather than passively preventing dense gas formation. CMR shows another facet of B-fields in the interstellar medium (ISM).

Before we model the formation of Orion A with CMR, there is still much to learn about the physical process. In particular, we still lack knowledge of how CMR reacts to different initial conditions. K21 only explored a limited number of cases for CMR. In the ISM, cloud-cloud collisions can be different from the specific condition for the Stick in K21. In this paper, to show the variation of CMR, we explore a number of different initial conditions for CMR for Stick-like filaments ($\sim$1 pc). The exploration prepares us for future modeling of the Orion A cloud and helps us understand the dynamical evolution of the ISM.

In the following section \S\ref{sec:method}, we briefly introduce the numerical method for the exploration. In \S\ref{sec:exploration}, we describe in detail the initial conditions for our explorations. In \S\ref{sec:results}, we report the exploration results and the analyses, followed by \S\ref{sec:discus} in which we discuss the implications. Finally, \S\ref{sec:summary} summarizes and concludes the paper.

\section{Method}\label{sec:method}

\subsection{Numerical Simulation}\label{subsec:sim}

Following K21, we use the public code Athena++ 
\citep{2020ApJS..249....4S} to model the CMR-filament. 
The code setup is basically the same as that in K21, 
i.e., we model the compressible, 
isothermal (with temperature $T$), inviscid
magnetohydrodynamics (MHD), with self-gravity
and Ohmic resistivity ($\eta$).
A uniform Cartesian grid is adopted, 
with periodic boundary conditions for all dimensions.
In each dimension, the grid has 512 cells and extends a physical
scale of 4 pc, each cell being 0.0078 pc (1600 AU).
The code unit for mass density is set to 
$3.84\times10^{-21}$ g cm$^{-3}$ ($n_{\rm H_2}$=840 cm$^{-3}$,
assuming a mean molecular mass per H$_2$ of 
$\mu_{\rm H_2}=2.8 m_\textrm{H}$).
The code unit for time is set to 2.0 Myr. 
The code unit for length scale is set to 1.0 pc. 
The code unit for velocity is set to 0.51 km s$^{-1}$. 
With these setup, the gravitational constant is $G=1$ in code unit. 
The B-field code unit is 3.1 $\mu$G.

\subsection{Synthetic Observation}\label{subsec:synth}

We utilize the radiative transfer (RT) code RADMC-3D \citep{2012ascl.soft02015D} to model the emission of CMR-filaments from the simulations. In this pilot study, we focus on the thermal dust emission because it has several advantages over the molecular line emission. First, a dust emission image in a single band is much easier to handle than a multi-channel line emission cube. It gives a quick overview of the target without taking too much space. Second, there are many large-scale dust emission surveys by, e.g., Herschel \citep{2010A&A...518L.102A,2016A&A...591A.149M} and APEX \citep{2009A&A...504..415S}. They provide a large pool of molecular clouds for our purpose of finding CMR-filaments. Therefore, in this paper, we only include results based on dust emission synthetic observations.

As shown in K21, the Stick filament was most prominent at far infrared wavelengths ($\lambda\ga100\mu$m). In particular, the Herschel 160/250 $\mu$m images showed the best balance between sensitivity and resolution. We thus focus on these two wavelengths in our modeling and use the Herschel data archive as our initial data pool. On the one hand, molecular clouds are typically best traced by cold dust emission at far infrared wavelengths, which is not affected by optical depth and depletion in molecular lines. On the other hand, we want to focus on the shorter wavelengths in far infrared to gain some spatial resolving power. The Herschel 160/250 $\mu$m band has a beam size of 12/18 arcsec which is better than, e.g., the ATLASGAL sub-mm survey (19 arcsec). At wavelengths shorter than 160 $\mu$m, the Stick filament in K21 became less prominent (see their figure 1 for the 100 $\mu$m panel). At 70 $\mu$m, the filament became optically thick and disappeared. Therefore, the 160 $\mu$m band is probably the shortest wavelength we can use. Finally, although powerful interferometers like ALMA can trace cloud details at mm wavelengths, they typically lose fluxes at certain spatial scales. They may be used to confirm CMR-filament candidates by tracing detailed structures, which will require the combination with compact arrays. 

Since our simulations are isothermal, we simply assume a same dust temperature as the gas temperature. For the fiducial model, the gas temperature $T$=15 K. So in RADMC-3D, the dust temperature is not computed by the RT code. We manually provide the dust temperature grid in "dust\_temperature.dat". The dust density distribution ("dust\_density.inp") is computed based on the gas density assuming a standard gas-to-dust mass ratio of 141. For the dust opacity, we use the "dustkappa\_silicate.inp" table from the RADMC-3D code. The table includes the mass-weighted absorption opacity $\kappa_{\rm abs}$ and the mass-weighted scattering opacity $\kappa_{\rm scat}$ as a function of wavelength from 0.1 $\mu$m to 10000 $\mu$m. The opacities are for an amorphous spherical silicate grain with a radius of 0.1 $\mu$m. The absorption opacity at 250 $\mu$m is about a factor of 4 smaller than the thin ice mantle opacity from \citet[][hereafter OH94]{1994A&A...291..943O} in which dust evolution was considered. Since our simulations are not tracking dust evolution, and a comparison between our RT images with the SimLine3D RT code \citep[with the OH94 dust model][]{Ossenkopf2002} shows no difference in filament morphology (besides a constant scaling factor of $\sim$4), we continue using the RADMC-3D dust opacity input because we focus on the filament morphology. A more appropriate dust model will be used when we compare with observations.

In observation, a filament is randomly oriented relative to our line-of-sight. So we randomly select the inclination angle, the azimuth angle, and the rotation angle for the RT model. Currently, we have 400 combinations of the angles for each time step of each exploration. In the results section, we will show 3 projections for each time step. The rest will be used for the statistical analysis in \S\ref{subsec:fil}.

\subsection{Filament Characterization}\label{subsec:fil}

\begin{figure*}[htb!]
\centering
\includegraphics[angle=0,width=\textwidth]{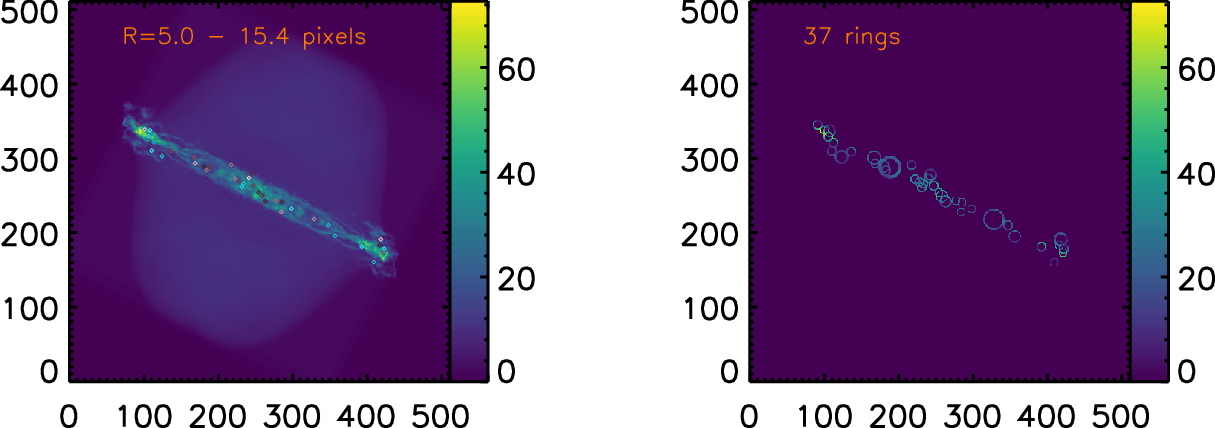}
\caption{
Illustration of the ring finding results for the projection of the fiducial model, MRCOL, that shows most ring and fork structures. The left panel shows the column density map where the centers of the 37 detected rings are marked by diamonds. The shade of grey of the diamond characterizes the contrast of the ring changing from 0.02 for almost black symbols to 0.4 for the strongest ring marked by a white diamond. The right panel visualizes the rings by showing only that fraction of the column density map that falls within the corresponding ring annuli.
\label{fig:rings}}
\end{figure*}

Visual inspection of the projected filament maps shows a large variation of filament widths and substructures as a function of projection (see \S\ref{sec:results}). The observed structure depends strongly on the orientation of the initial cloud collision to the line of sight, a parameter that is a priori not known to an observer. For a comparison to observational data it is therefore important to get the statistics of filament parameters over all possible angles to judge the likelihood of a match between a particular model and observed data.

Here we use two statistical measures to characterize the filaments. The first is the power of anisotropic fluctuations in the maps as a function of their width, basically measuring the width of the dominating structure and characterizing how this main filament is composed from contributions of subfilaments at smaller scales. The second is a "ringiness" of the filament given by the number of ring-like substructures and their contrast to the environment. It measures to what degree the filament consists of rings and forks seen as a possible trace of CMR in the Stick filament (see Sect.~\ref{sec:results}).

The width analysis performs a wavelet decomposition of the map using filament-like anisotropic wavelets. The power of the wavelet spectrum as a function of the wavelet width then shows to what degree the map consists of elongated structures with that width \citep{OssenkopfOkadaStepanov2019}. In this way, the tool looks for any filamentary structure so that it detects a clear contribution at the width of the main filament, but it also finds the threads within the main filaments. The statistically dominant width is not necessarily the width of the main filament but can be smaller. We compute the full spectrum of wavelet coefficients to measure the contribution of different widths to the overall structure. As the method can be applied to any map, it is useful for a general comparison to the observations \citep{OssenkopfOkadaStepanov2019}. 

The characterization of rings in the map follows a similar principle but implements a number of additional criteria. The map is also filtered with a wavelet, this time a radially symmetric ring, that is positive in an annulus and slightly negative in the core so that the total integral is zero. For the CMR analysis we restricted the ratio between the outer and the inner radius of the annulus to a range of 1.2-1.4, matching typical eye assignments of ring structures. We then searched the wavelet convolved map, representing the covariance between the original map and the ring wavelet for peaks. To identify statistically significant peaks within the map we used the established GAUSSCLUMPS algorithm \citep{StutzkiGuesten1990} but the choice of the exact peak finding algorithm has almost no impact on the result.

The covariance map peaks at the points with the strongest match between the actual structure and the ring. However, it also peaks at other structures providing some match with the ring annulus such as simple point sources or sharp edges. Therefore we filter the peaks by three additional criteria. a) To exclude the detection of curved edges we require that the peaks are relatively isotropic with an aspect ratio below 2.0. b) To exclude the detection of small ring fragments and point sources we require a relatively uniform distribution of original map values within the detected ring candidate by setting an upper limit to the skewness of their distribution of 0.2. c) Finally to distinguish rings from uniformly filled circles we define a \textit{contrast} as the ratio of the wavelet coefficients for the ring-filtered map relative to the map that is filtered by filled circles of the same outer radius. Peaks where this contrast does not exceed 1.05 are ignored.

We also use this \textit{contrast} to quantify the significance of the rings. In total we characterize the "ringiness" in the map by the number of detected rings times their \textit{contrast}. This gives one value for each ring radius and each ratio between outer and inner radius providing a two-dimensional spectrum. If rings are not perfectly circular or somewhat extended they can be detected for multiple parameter combinations. For rings with centres falling into other detected rings we only count the ring with the highest amplitude contrast. To condense the results into a single number per map we have summed up all ring contributions for inner radii between 5 and 13 pixels and a radius ratio between 1.2 and 1.4. 

To illustrate the approach, we show in Figure~\ref{fig:rings} the result for the projection of the fiducial model that contains most rings. One sees already by eye that the filament breaks up into many substructures through many bifurcations that are inhomogeneously distributed along the filament. The algorithm detects rings covering almost the whole radius range that is scanned because of the hierarchical breakup of the structure. By comparing both panels of the figure one can see what type of ring is detected at what position. With 37 rings and an average contrast of 0.11, the example shown here represents the upper extreme of the ``ringiness'' distribution for all projections shown in Figure~\ref{fig:allrings}.

\section{Initial Conditions}\label{sec:exploration}

\begin{deluxetable*}{@{\extracolsep{4pt}}cccccccccccccc}[htb!]
\tablecaption{Model Initial Conditions \label{tab:ic}}
\tablehead{
\colhead{Model} &
\colhead{\#} &
\colhead{$T$} &
\colhead{$\eta$} &
\multicolumn{5}{c}{Cloud1} &
\multicolumn{5}{c}{Cloud2} \\
\cline{5-9}
\cline{10-14}
\colhead{} &
\colhead{} &
\colhead{} &
\colhead{} & 
\colhead{$\rho_1$} & 
\colhead{$R_1$} &
\colhead{$v_{\rm 1,x}$} & 
\colhead{$v_{\rm 1,z}$} &  
\colhead{$B_{\rm 1,z}$} &
\colhead{$\rho_2$} & 
\colhead{$R_2$} & 
\colhead{$v_{\rm 2,x}$} & 
\colhead{$v_{\rm 2,z}$} & 
\colhead{$B_{\rm 2,z}$}
}
\startdata
\mbox{MRCOL} (\S\ref{sec:intro}) & (0) & 15 & 0.001 & 0.5 & 0.9 & 2.0 & 0.5 & 3.2 & 0.5 & 0.9 & -2.0 & -0.5 & -3.2 \\
\mbox{$\eta$\_L} (\S\ref{subsec:resist}) & (1) & 15 & \underline{0.0001} & 0.5 & 0.9 & 2.0 & 0.5 & 3.2 & 0.5 & 0.9 & -2.0 & -0.5 & -3.2 \\
\mbox{$\eta$\_H} (\S\ref{subsec:resist}) & (2) & 15 & \underline{0.01} & 0.5 & 0.9 & 2.0 & 0.5 & 3.2 & 0.5 & 0.9 & -2.0 & -0.5 & -3.2 \\
\mbox{$B$\_L} (\S\ref{subsec:b}) & (3) & 15 & 0.001 & 0.5 & 0.9 & 2.0 & 0.5 & \underline{1.6} & 0.5 & 0.9 & -2.0 & -0.5 & \underline{-1.6} \\
\mbox{$B$\_H} (\S\ref{subsec:b}) & (4) & 15 & 0.001 & 0.5 & 0.9 & 2.0 & 0.5 & \underline{6.4} & 0.5 & 0.9 & -2.0 & -0.5 & \underline{-6.4} \\
\mbox{$\rho_2$\_L} (\S\ref{subsec:dens}) & (5) & 15 & 0.001 & 0.5 & 0.9 & 2.0 & 0.5 & 3.2 & \underline{0.25} & 0.9 & -2.0 & -0.5 & -3.2 \\
\mbox{$\rho_2$\_H} (\S\ref{subsec:dens}) & (6) & 15 & 0.001 & 0.5 & 0.9 & 2.0 & 0.5 & 3.2 & \underline{1.0} & 0.9 & -2.0 & -0.5 & -3.2 \\
\mbox{$R_2$\_L} (\S\ref{subsec:radii}) & (7) & 15 & 0.001 & 0.5 & 0.9 & 2.0 & 0.5 & 3.2 & 0.5 & \underline{0.45} & -2.0 & -0.5 & -3.2\\
\mbox{$R_2$\_H} (\S\ref{subsec:radii}) & (8) & 15 & 0.001 & 0.5 & 0.9 & 2.0 & 0.5 & 3.2 & 0.5 & \underline{1.8} & -2.0 & -0.5 & -3.2 \\
\mbox{$T$\_L} (\S\ref{subsec:T}) & (9) & \underline{10} & 0.001 & 0.5 & 0.9 & 2.0 & 0.5 & 3.2 & 0.5 & 0.9 & -2.0 & -0.5 & -3.2\\
\mbox{$T$\_H} (\S\ref{subsec:T}) & (10) & \underline{30} & 0.001 & 0.5 & 0.9 & 2.0 & 0.5 & 3.2 & 0.5 & 0.9 & -2.0 & -0.5 & -3.2 \\
\mbox{$v_x$\_L} (\S\ref{subsec:vcol}) & (11) & 15 & 0.001 & 0.5 & 0.9 & \underline{1.0} & 0.5 & 3.2 & 0.5 & 0.9 & \underline{-1.0} & -0.5 & -3.2 \\
\mbox{$v_x$\_H} (\S\ref{subsec:vcol}) & (12) & 15 & 0.001 & 0.5 & 0.9 & \underline{4.0} & 0.5 & 3.2 & 0.5 & 0.9 & \underline{-4.0} & -0.5 & -3.2 \\
\mbox{$v_z$\_L} (\S\ref{subsec:vshe}) & (13) & 15 & 0.001 & 0.5 & 0.9 & 2.0 & \underline{0.25} & 3.2 & 0.5 & 0.9 & -2.0 & \underline{-0.25} & -3.2 \\
\mbox{$v_z$\_H} (\S\ref{subsec:vshe}) & (14) & 15 & 0.001 & 0.5 & 0.9 & 2.0 & \underline{1.0} & 3.2 & 0.5 & 0.9 & -2.0 & \underline{-1.0} & -3.2
\enddata
\tablecomments{$T$ is the isothermal temperature. $\eta$ is the Ohmic resistivity. All cloud parameters are defined in Figure \ref{fig:cmr}. All numeric values are in code unit (see \S\ref{sec:method}). The first model \mbox{MRCOL} is the fiducial model from K21. The underline marks the explored parameter.}
\end{deluxetable*}

The setup for the cloud-cloud collision is the same as K21.
Figure \ref{fig:cmr} shows the initial condition for the
fiducial model \mbox{MRCOL} in K21.
The two clouds, moving along the x-axis, collide at the origin.
Cloud1 has density $\rho_1=0.5$, radius $R_1=0.9$, colliding velocity
$v_{\rm 1,x}=2.0$, shear velocity $v_{\rm 1,z}=0.5$,
B-fields $B_{\rm 1,z}=3.2$. 
Cloud2 has $\rho_2=0.5$, radius $R_2=0.9$, colliding velocity
$v_{2,x}=-2.0$, shear velocity $v_{\rm 2,z}=-0.5$,
B-fields $B_{\rm 2,z}=-3.2$. 
These fiducial parameter values are listed in the 
first row of Table \ref{tab:ic} (model \#0). 

Starting from the second row in Table \ref{tab:ic},
we present 14 models, exploring 7 parameters,
including the Ohmic resistivity (models \#1 and \#2), 
the B-field (models \#3 and \#4), 
the Cloud2 density (models \#5 and \#6), 
the Cloud2 radius (models \#7 and \#8), 
the temperature (models \#9 and \#10),
the collision velocity (models \#11 and \#12),
the shear velocity (models \#13 and \#14). 
For each parameter, we explore a lower value (marked with "L"), 
and a higher value (marked with "H").

In the first part of the results section \S\ref{sec:results},
we mainly consider three properties of the filament, including 
the density probability distribution function ($\rho$-PDF),
the morphology, and the relation between B-magnitude and 
density ($B$-$\rho$). As shown in K21, the CMR-filament has rich
sub-structures, including spikes, rings, and forks.
They form due to the dynamical effect of B-fields.
Showing the filament morphology provides a first impression
of the filament structure. The fiducial CMR-filament
in K21 was ruler-straight, matching the Stick filament.
However, as the explorations will show, CMR-filaments 
can be curved in certain conditions. In the second part
of the results section, we quantify the filament width
and the ringiness in the filament, utilizing the
methods introduced in \S\ref{subsec:fil}.

\section{Results and Analysis}\label{sec:results}

In the following, we show a multipanel figure for each
exploration at different time steps, including the filament
density distribution, the dust emission at 250 $\mu$m 
(the 160 $\mu$m emission is very similar), 
and the $B$-$\rho$ relation where $B$ is the B-field magnitude.
The statistics are computed within the central 1.6 pc region.
Only the figure for the fiducial model below includes t=0.

\begin{figure*}[htb!]
\centering
\epsscale{1.15}
\plotone{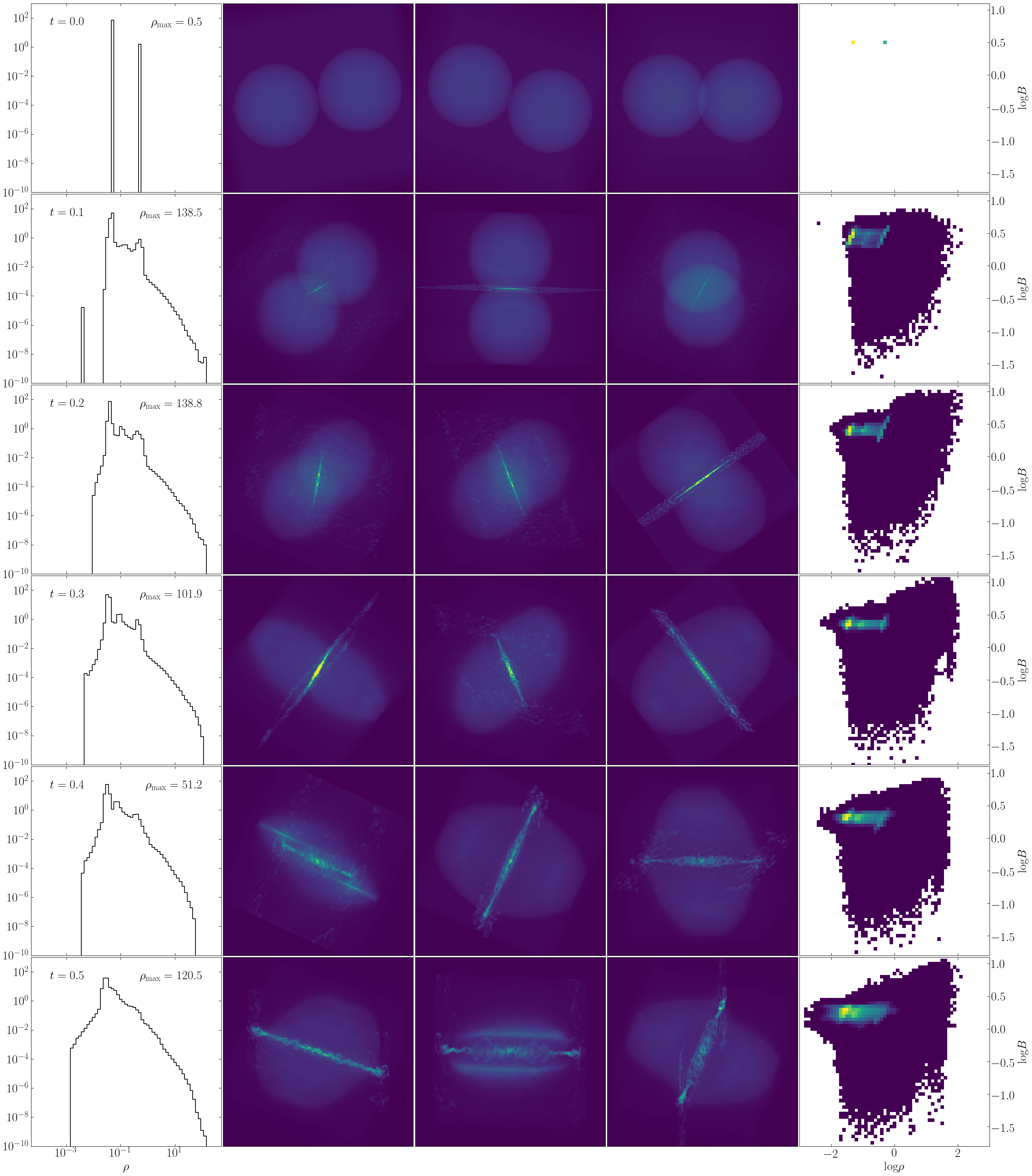}
\caption{
\mbox{MRCOL} results (model \#0 in Table \ref{tab:ic}). Each row shows the results from a time step labeled in the leftmost panel. The first column shows the $\rho$-PDF from $5\times10^{-5}$ to $5\times10^2$. The middle three columns show the RT images at different projections. They have the same color scale (linear) and range (0-100 MJy sr$^{-1}$). No noise is implemented. The fifth column shows the $B$-$\rho$ relation as a 2D histogram. $\rho$ ranges from $10^{-3}$ to $10^3$ and $B$ ranges from 0.014 to 15. The color scale is in probability density in logarithmic scale (10$^{-2}$ to 10$^{2}$).
\label{fig:mrcol}}
\end{figure*}

Figure \ref{fig:mrcol} shows the fiducial model result from K21. It is set as a reference for comparison with other exploration results. We are focusing on the early stages from t=0 to t=0.5 (1 Myr) of the filament formation for three reasons. First, it is a natural follow-up of the K21 study in which they compare the t=0.3 (0.6 Myr) results with the Stick filament which is a very young filament. Here we include time steps up to 1 Myr to give more information. Second, due to the periodic boundary condition, MHD waves will propagate back in the computation domain at $t\ga1$ Myr. Third, the filament starts to collapse after 1 Myr and star formation happens \citep{2022MNRAS.517.4679K}. We limit our study to the starless phase and focus on filament properties.

The first column in Figure \ref{fig:mrcol} shows the $\rho$-PDF for each time step. Initially, there were just two densities, i.e., the cloud density of 0.5 ($n_{\rm H_2}=420$ cm$^{-3}$) and the ambient density of 0.05 ($n_{\rm H_2}=42$ cm$^{-3}$). At t=0.1, the density distribution broadens significantly, with the maximum value reaching 138.5 ($n_{\rm H_2}=1.2\times10^5$ cm$^{-3}$). Using the VisIt tool \citep[][chapter 16]{HPV:VisIt}, we are able to locate the high-density cells in the 3D domain\footnote{The tool is publicly available at \url{https://visit-dav.github.io/visit-website/}. It can read in the Athena++ xdmf file which is described at \url{https://github.com/PrincetonUniversity/athena/wiki}. The exploration data is available in the Harvard Dataverse at \url{https://doi.org/10.7910/DVN/CXHWRR}.}. They are near the center of the filament.

\begin{figure*}[htb!]
\centering
\epsscale{1.1}
\plottwo{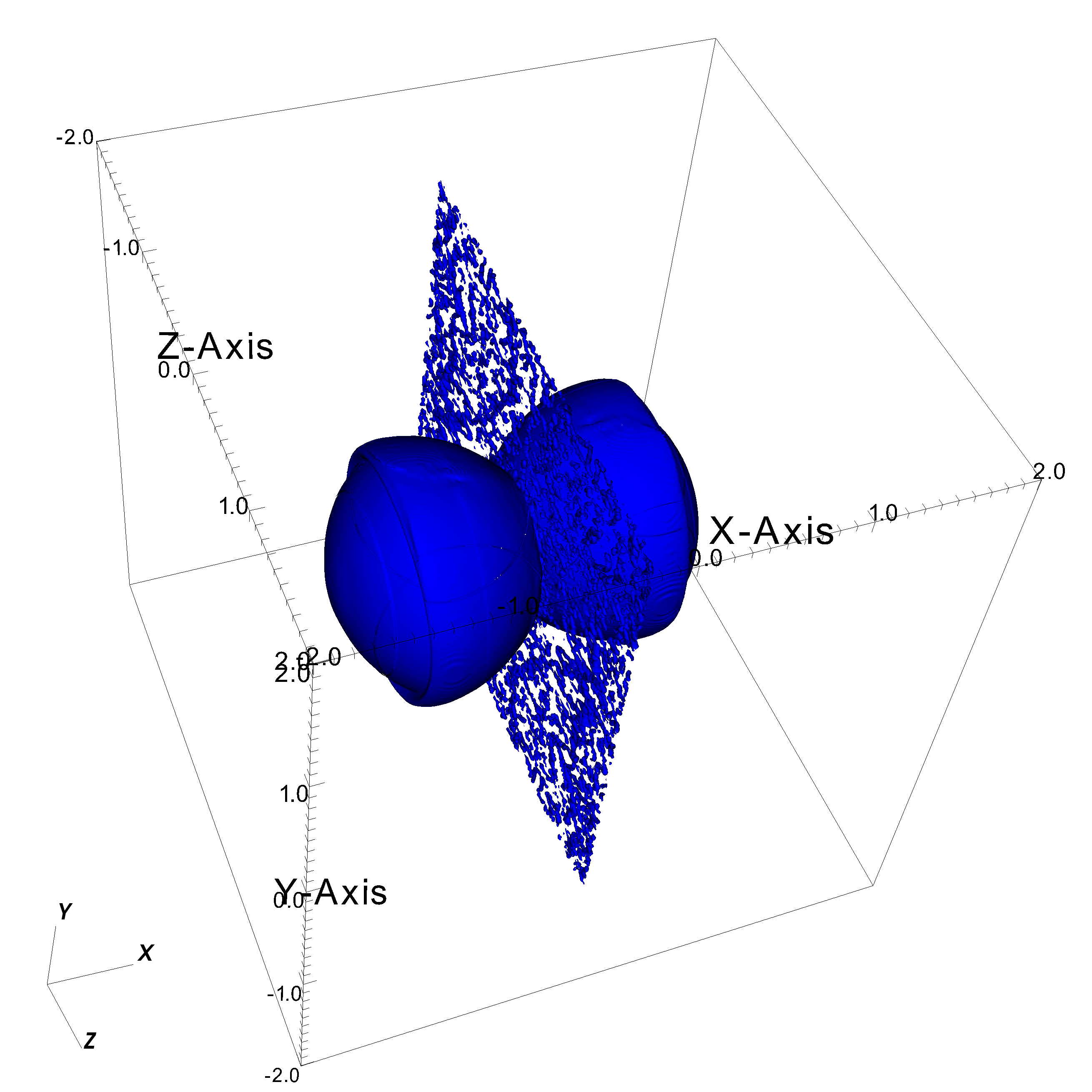}{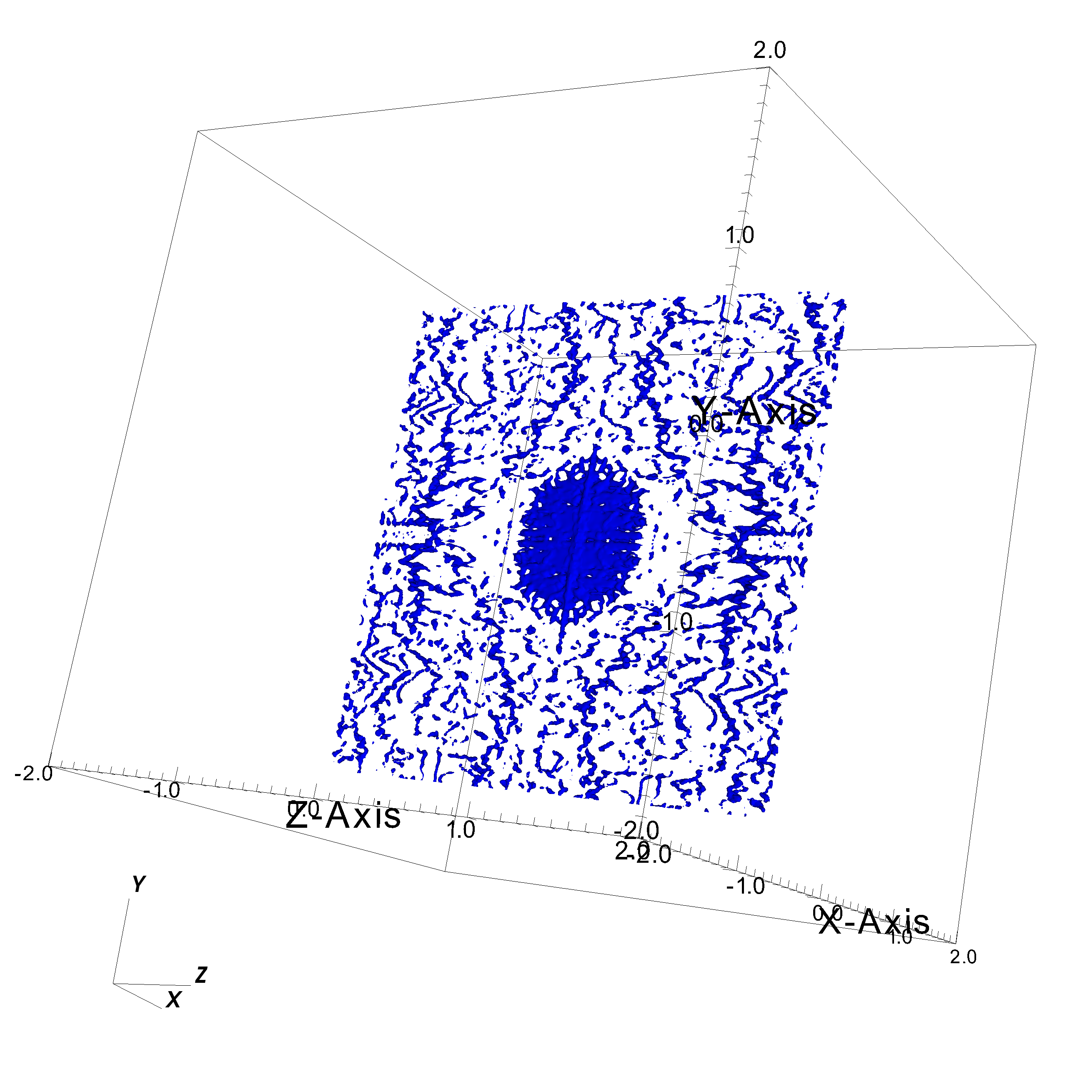}\\
\plottwo{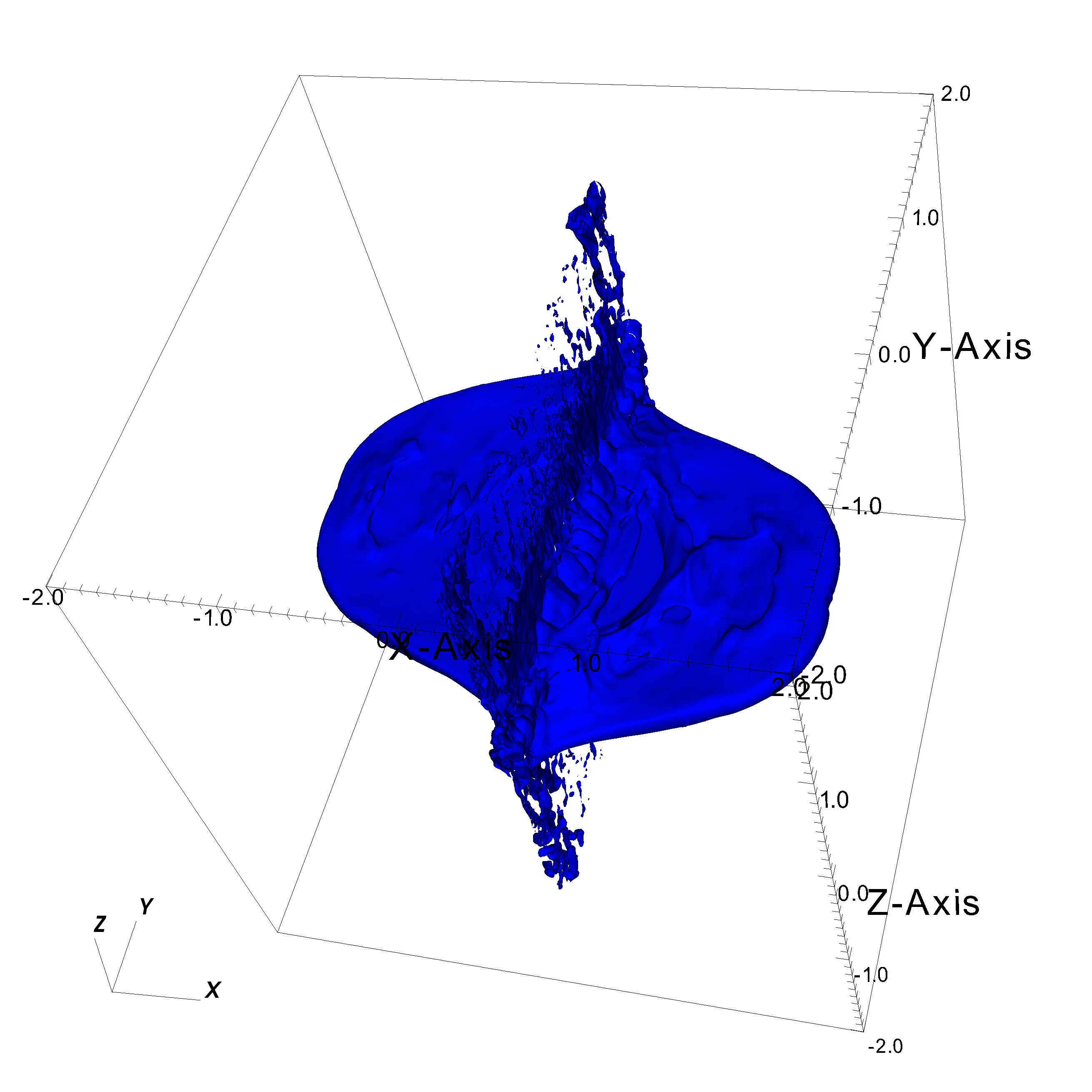}{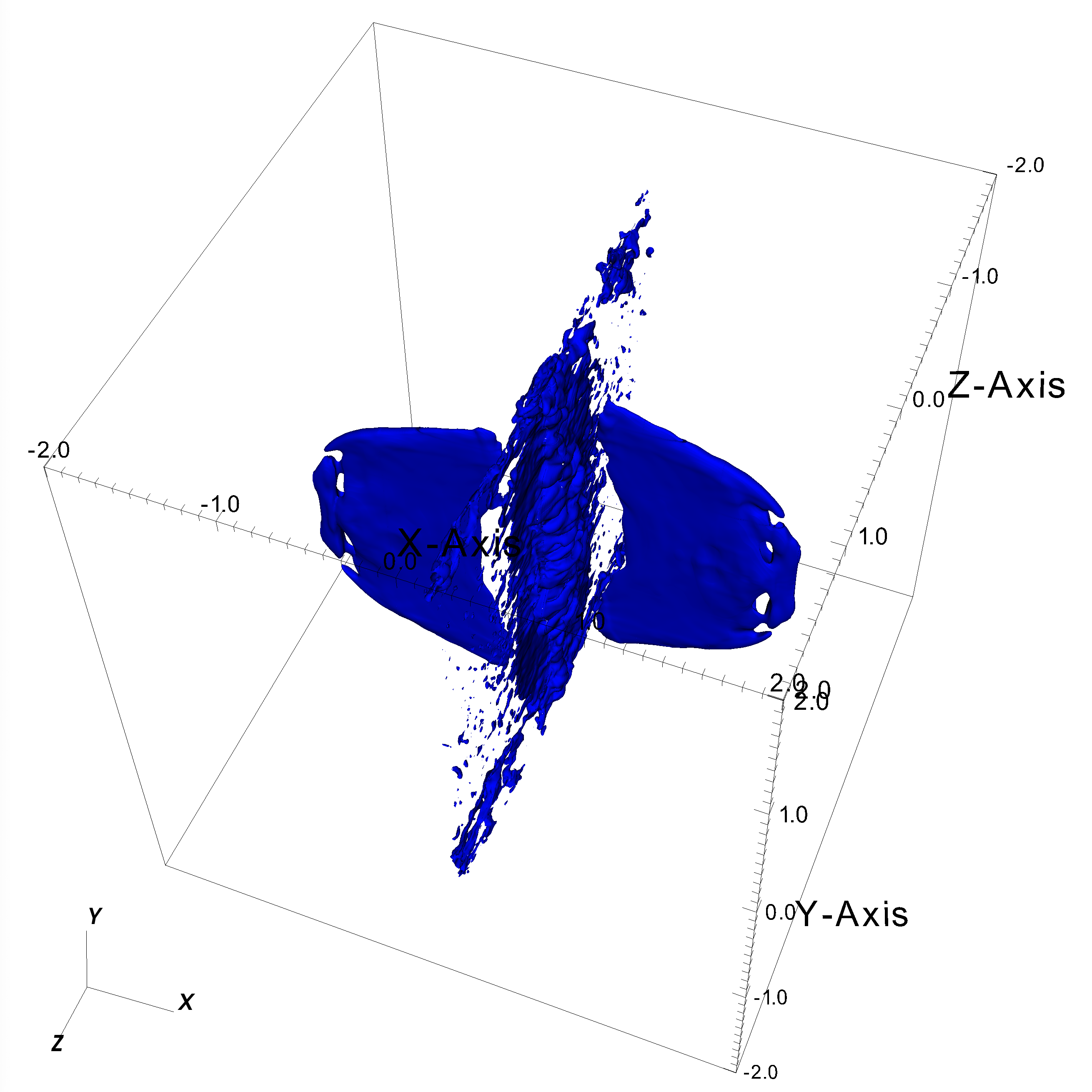}
\caption{
Top row: isosurface of density at t=0.1 in \mbox{MRCOL}. The left panel shows $\rho=0.5$ and the right panel shows $\rho=1$. Bottom row: isosurface of density at t=0.5 in \mbox{MRCOL}. The left panel shows $\rho=0.5$ and the right panel shows $\rho=1$.
\label{fig:mrcolrho}}
\end{figure*}

Using the isosurface slicing, we see that cells with $\rho\ga5$ are preferentially in the filament. Lower density cells ($\rho\ga0.7$) are in the field-reversing plane (the collision midplane). Figure \ref{fig:mrcolrho} top row shows two density isosurfaces from the t=0.1 output using the VisIt tool. On the left we see $\rho=0.5$ which is the initial cloud density. Due to reconnection, the field-reversing midplane forms gas fibers at a variety of  densities. They constitute the dense gas in the $\rho$-PDF at t=0.1 in Figure \ref{fig:mrcol}. The right panel shows $\rho=1.0$ gas which is solely in the midplane. The most important structure is the pancake and the filament cutting through the middle of the pancake.

With the progress of the simulation, the $\rho$-PDF continues to broaden into the low-density regime. At t=0.5, the lowest density reaches $\sim10^{-3}$ ($n_{\rm H_2}=0.84$ cm$^{-3}$). Again, using the VisIt tool, we find that the two clouds collapse along the z-axis (the direction of the field lines) and form two flat structures (see Figure \ref{fig:mrcolrho} bottom row). The highest density in the flat structure reaches $\sim$2. They show no signs of fragmentation. Due to the collapse, the flat structures develop a density gradient along the z-direction. The low-density cells in the PDF are on the outer shells. Further out, the gas structure is slightly affected by the periodic boundary condition (even though we exclude the edges), so the low-density part of the PDF could be artifact. Meanwhile, the maximum density remains $\sim$138 at t=0.2 but drops to 51 at t=0.4. At t=0.5, a high density tail $\ga50$ builds up again, with the maximum density reaching 120. They are at the two ends of the filament. Notably, cells with $\rho\ga5$ are preferentially in the filament throughout the simulation.

\begin{figure*}[htb!]
\centering
\epsscale{1.1}
\plottwo{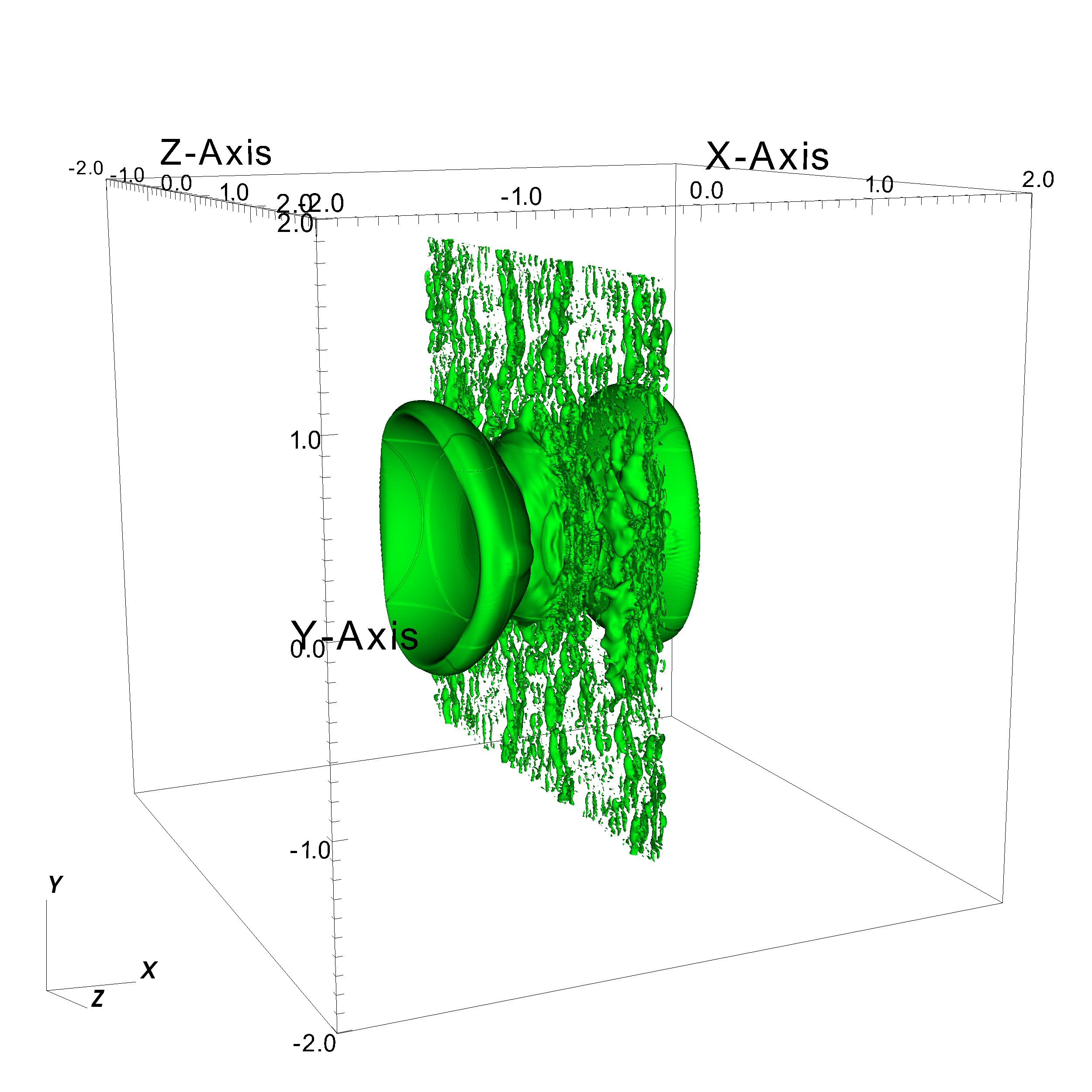}{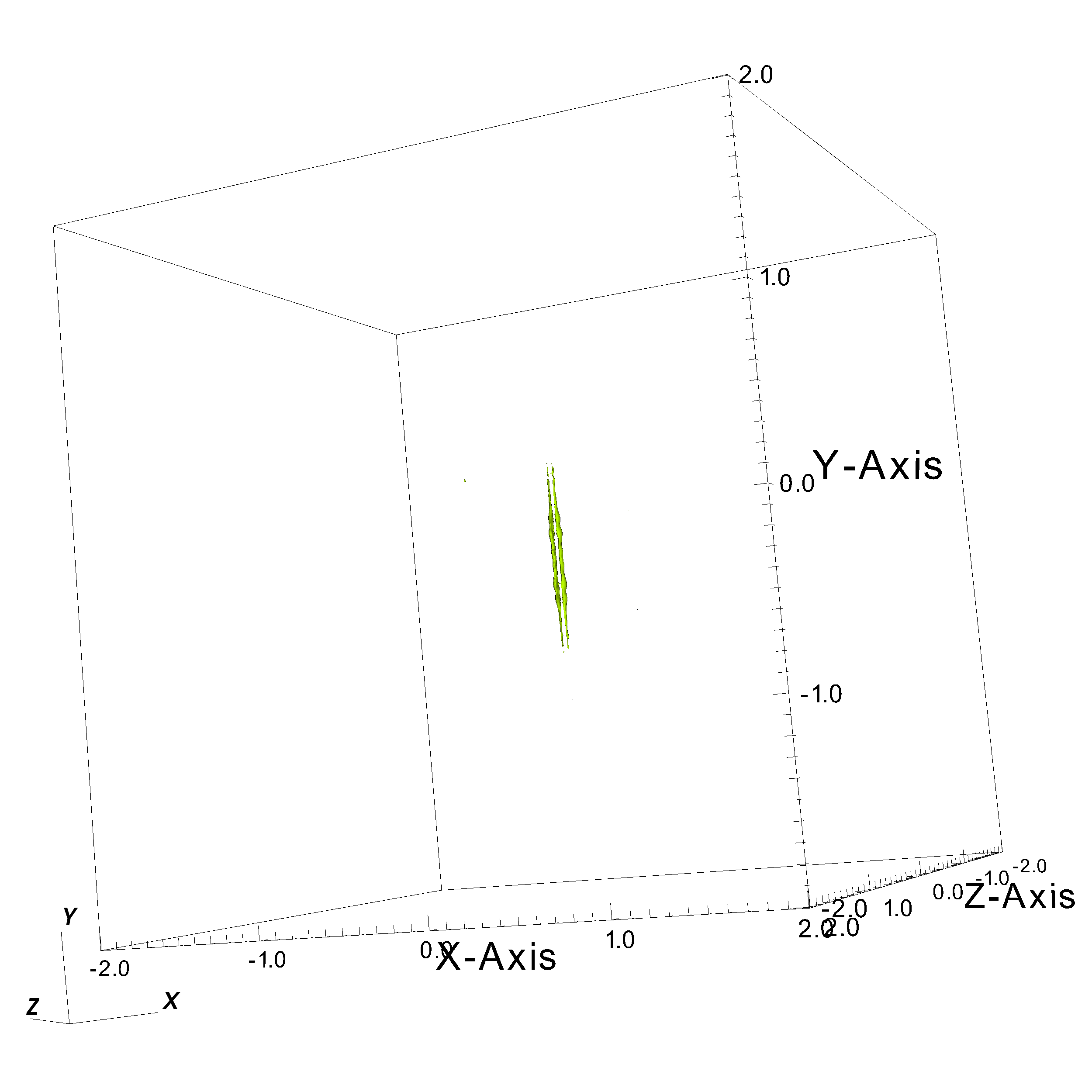}\\
\plottwo{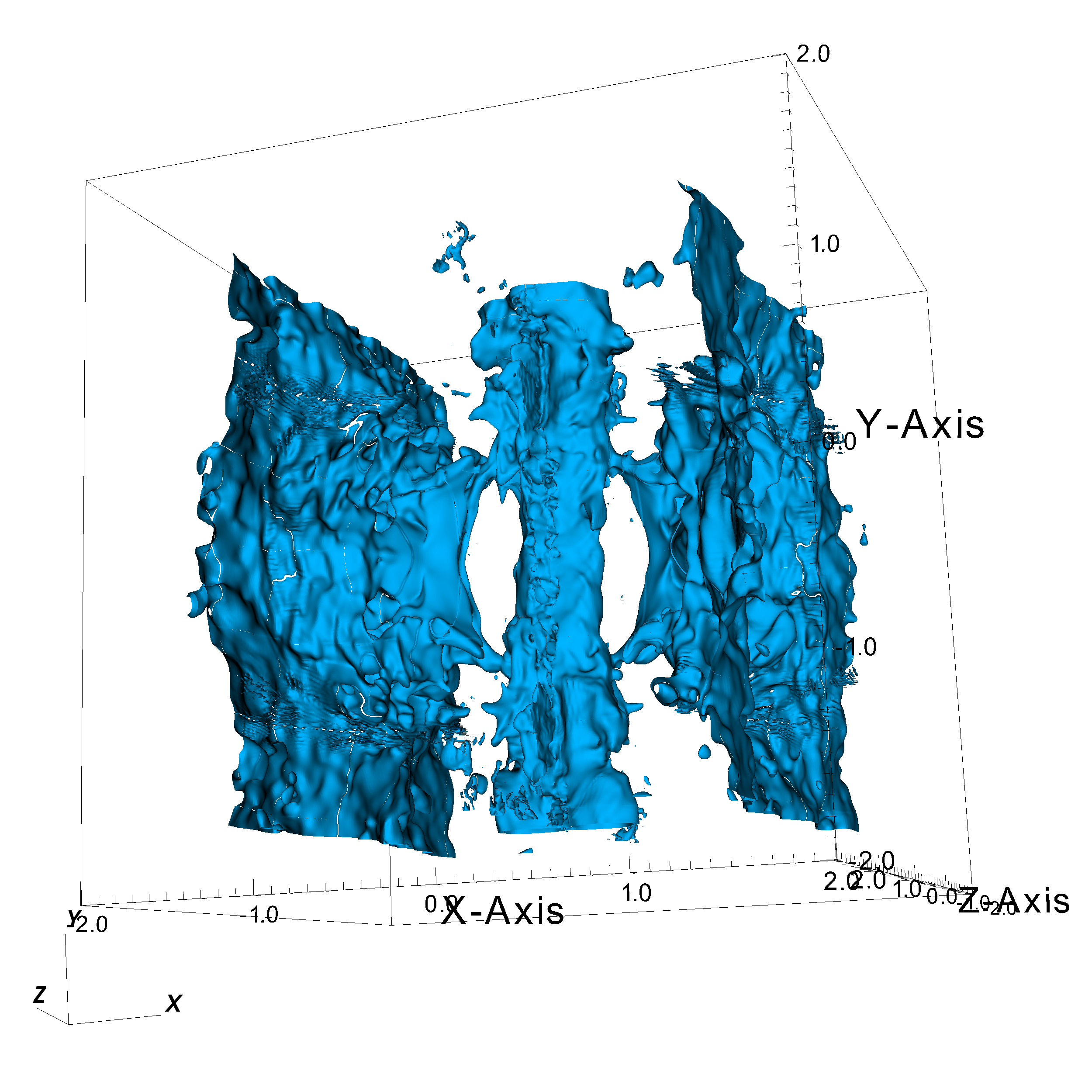}{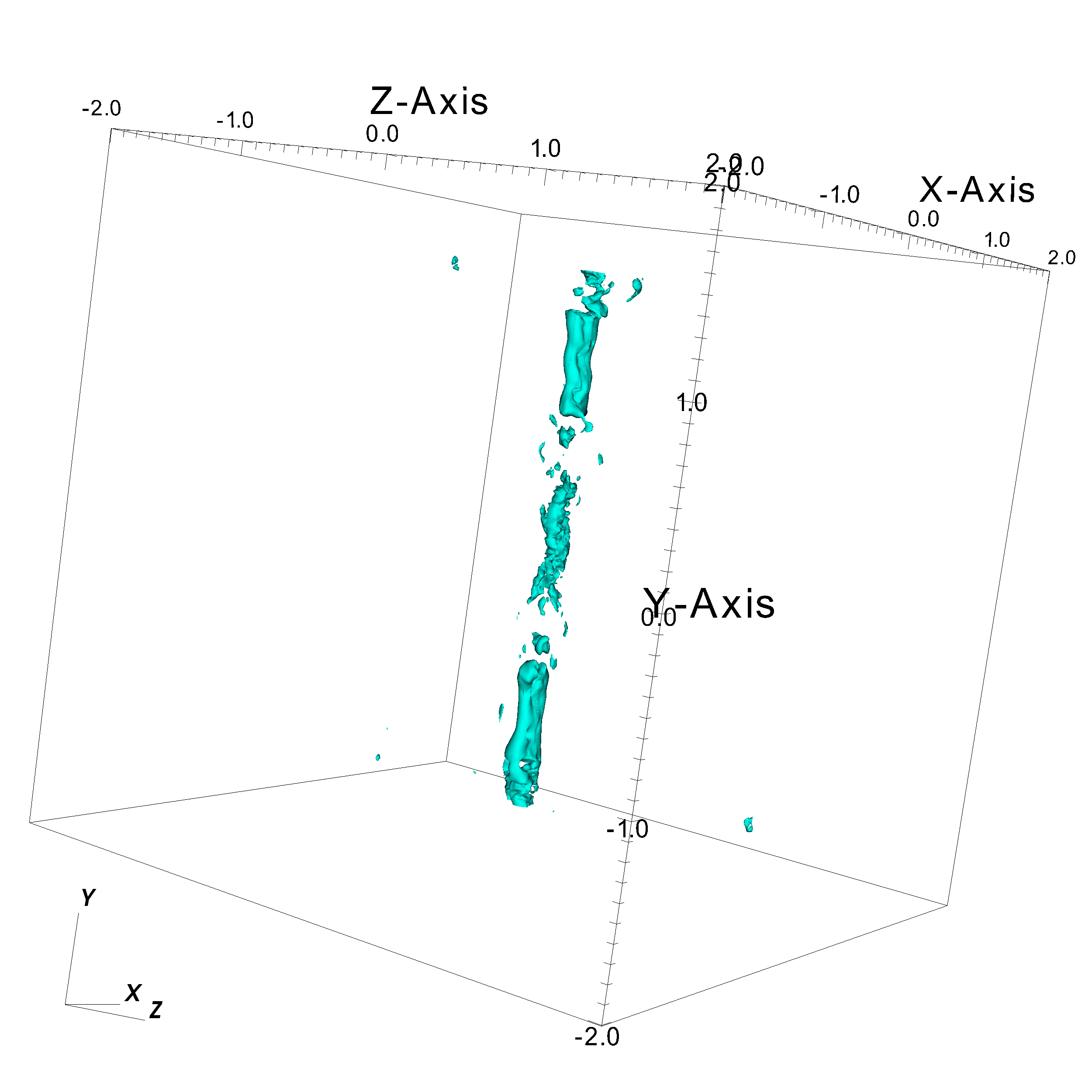}
\caption{
Top row: isosurface of B-magnitude at t=0.1 in \mbox{MRCOL}. The left panel shows $B=3.5$ and the right panel shows $B=5$. Bottom row: isosurface of B-magnitude at t=0.5 in \mbox{MRCOL}. The left panel shows $B=2$ and the right panel shows $B=3$. 
\label{fig:mrcolB}}
\end{figure*}

The second to fourth columns in Figure \ref{fig:mrcol} show the synthetic observations of the 250 $\mu$m dust emission (see \S\ref{subsec:synth}). The first row shows the initial two clouds. The second row shows the collision and the formation of the filament in the collision midplane. The filament grows longer and thicker over time. Rings and forks show up along the filament due to the projection of many randomly-oriented, curved fibers. For example, at t=0.5, the second projection (third column) shows the curly fibers. But in the first and third projections (second and fourth columns), the filament shows more ring-like sub-structures. This comparison shows that the ring count depends on the projection, which is why we take the average over all projections (\S\ref{subsec:fil}).

The fifth column in Figure \ref{fig:mrcol} shows the $B$-$\rho$ relation at each step in 2D histograms. Initially, the computation domain only has one value of B-field $B=3.2$ (10 $\mu$G) and two values of density (the cloud density and the ambient density). With the progress of the simulation, the $B$-$\rho$ plane is quickly populated (second to sixth rows) with a wide range of density and B-field magnitudes. The initial two points extend to a blob of bright pixels in the $B$-$\rho$ plot. 

At t=0.1, cells with $B>3.2$ (the initial field magnitude) concentrate on two sides of the collision midplane that are immediately contacting the filament (Figure \ref{fig:mrcolB} top-left), including the cloud surface that is pointing toward the filament. This field elevation is due to the compression effect from the collision and is responsible for confining the filament. Figure \ref{fig:mrcolB} top-right shows the confinement better. One can see two parallel elongated structures with $B=5$ that are immediately contacting the two sides of the filament. Their higher B-magnitude means higher magnetic pressure $B^2/8\pi$, which confines the filament in the x-direction. From K21 we know that magnetic tension confines the filament in the z-direction. Together, these two forces maintain the filament structure before gravity dominates.

Cells with $B\lesssim3$ are spread over the outer shells of the clouds plus plane-parallel waves propagating away from the collision midplane (the field-reversing interface at x=0). On the other hand, the midplane contains both low and high B-magnitudes, except that highest B-magnitudes concentrate around the filament. See Figure \ref{fig:mrcolB} bottom row for the B-magnitude distribution at t=0.5. Note, at this point, gravity is not dominating yet. So the B-field distribution is not a result of gravitational collapse with flux-freezing which gives the power-law relation $B\propto\rho^{2/3}$.

In Figure \ref{fig:mrcol} at t=0.5, the dominating $B-\rho$ relation (the bright part of the histogram) is still flat. The majority of B-magnitudes is now at $\sim1.7$ (smaller than the initial value of 3.2). Figure \ref{fig:mrcolB} bottom-left shows the B-magnitude isosurface at $B=2$. On the other hand, cells with $B\ga4$ are preferentially in the filament throughout the simulation. Figure \ref{fig:mrcolB} bottom-right shows the B-magnitude isosurface at $B=3$. We can see that strong fields concentrate around the filament. Below, we show the results for each exploration parameter. Only major difference from the fiducial model \mbox{MRCOL} will be discussed. 

Based on the above analysis of the density and B-field, we can see that there are two distinct parts of the computation domain that evolve separately. One is the field-reversal interface where magnetic reconnection produces gas with a wide range of physical states. The other is the region outside the interface. Here the gas gradually evolves and deviates from the initial state due to gravity and the MHD without reconnection. 

\subsection{Resistivity}\label{subsec:resist}

\begin{figure*}
\centering
\epsscale{1.15}
\plotone{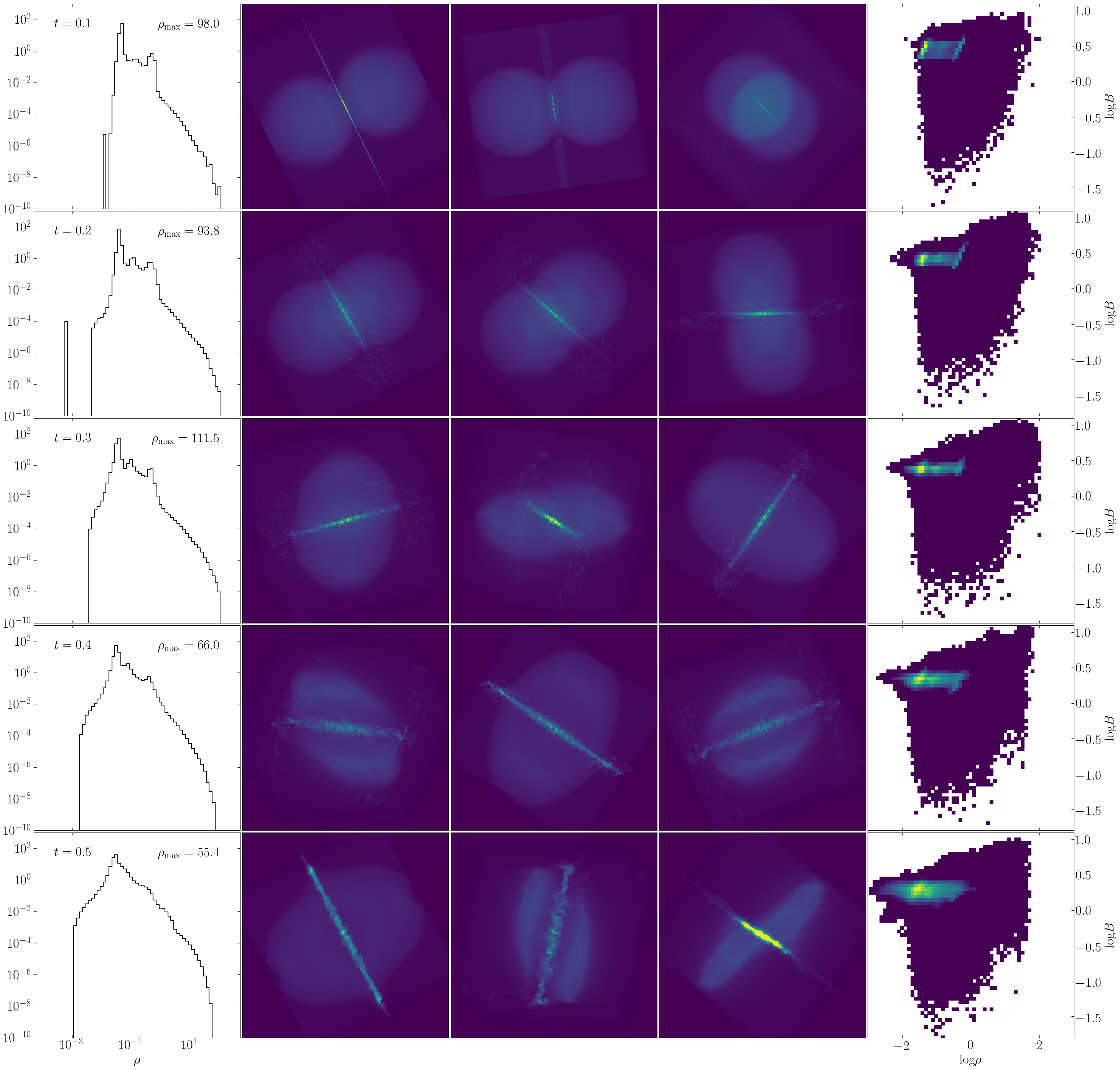}
\caption{
Same format as Figure \ref{fig:mrcol}, but for the \mbox{$\eta$\_L} model (\#1 in Table \ref{tab:ic}). 
\label{fig:re0p0001}}
\end{figure*}

\begin{figure*}
\centering
\epsscale{1.15}
\plotone{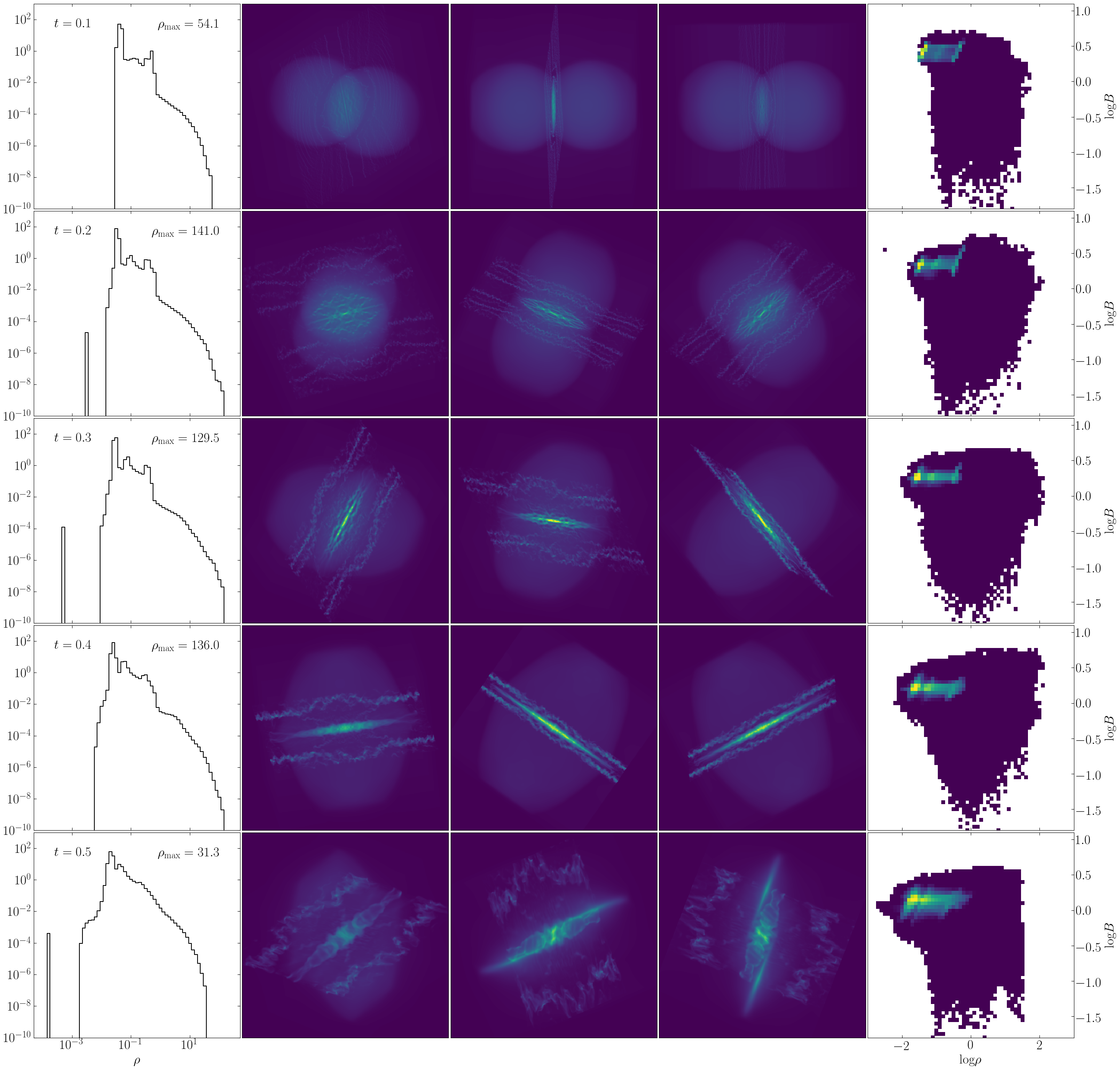}
\caption{
Same format as Figure \ref{fig:mrcol}, but for the \mbox{$\eta$\_H} model (\#2 in Table \ref{tab:ic}). 
\label{fig:re0p01}}
\end{figure*}

Figures \ref{fig:re0p0001} and \ref{fig:re0p01} show the results from \mbox{$\eta$\_L} and \mbox{$\eta$\_H}, respectively. The former has an Ohmic resistivity a factor of 10 smaller than \mbox{MRCOL}, which is close to the numerical resistivity ($\la$10$^{-4}$, K21). The latter has an Ohmic resistivity a factor of 10 larger than \mbox{MRCOL}.

Compared to \mbox{MRCOL}, \mbox{$\eta$\_L} has a lower maximum density from t=0.1 to t=0.2. From t=0.3 to t=0.4, \mbox{$\eta$\_L} has a slightly higher maximum density than \mbox{MRCOL}. At t=0.5, \mbox{$\eta$\_L} does not build up the high-density tail as that in \mbox{MRCOL}. On the other hand, \mbox{$\eta$\_H} has a lower maximum density than \mbox{MRCOL} at t=0.1. But from t=0.2 to t=0.4, the maximum density in \mbox{$\eta$\_H} is higher than \mbox{MRCOL}. At t=0.5, there is a cutoff at high densities in \mbox{$\eta$\_H}. Overall, the $\rho$-PDFs show similar shapes, although the maximum density fluctuates with time. This behavior indicates the unstable nature of magnetic reconnection.

Regarding the morphology, \mbox{$\eta$\_L} is not that different from \mbox{MRCOL}. However, \mbox{$\eta$\_H} is noticeably different from \mbox{MRCOL} in that \mbox{$\eta$\_H} forms several large-scale filaments on two sides of the main filament in the field-reversing plane. The main filament also appears to be smoother with less fibers, especially at later times. At t=0.5, the main filament becomes larger in diameter. 

In the $B$-$\rho$ relation, \mbox{$\eta$\_H} especially lacks cells with high B-magnitudes, which is consistent with the higher diffusion. In turn, the relatively weaker B-field at later times fails to hold dense gas, especially at t=0.5 when there is a cutoff at high densities. The above comparisons imply that relatively stronger B-field is necessary for CMR to maintain denser gas until gravity dominates. In fact, as shown in \citet{2022ApJ...933...40K}, what really matters is the plasma-$\beta$, i.e., the thermal-to-magnetic pressure ratio. If the thermal pressure is too high, the dense gas formation is suppressed. 

\subsection{Field Strength}\label{subsec:b}

\begin{figure*}
\centering
\epsscale{1.15}
\plotone{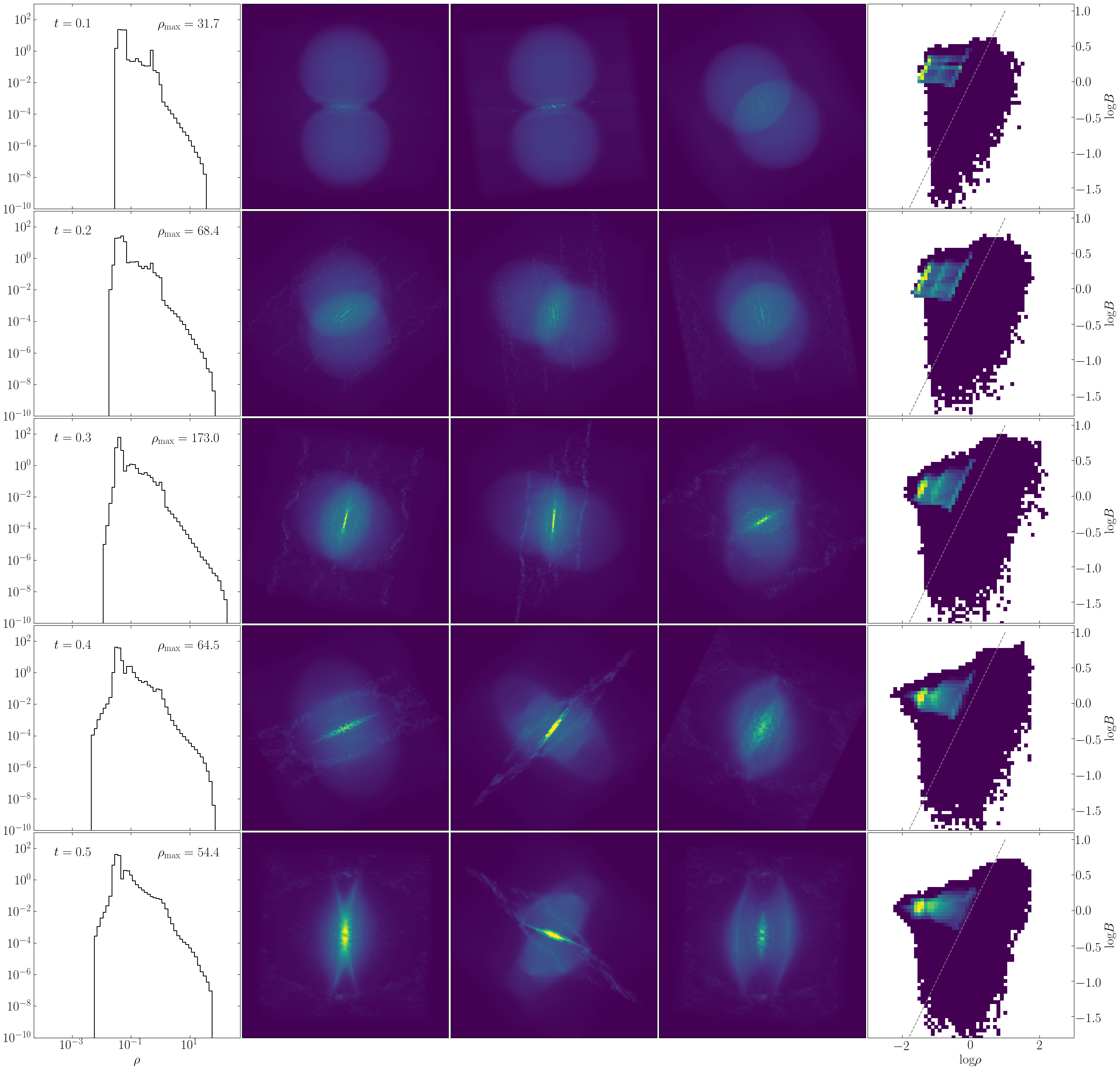}
\caption{
Same format as Figure \ref{fig:mrcol}, but for the \mbox{$B$\_L} model (\#3 in Table \ref{tab:ic}). The gray dashed line represents $B\propto\rho$ normalized at (0,0).
\label{fig:b1p5}}
\end{figure*}

\begin{figure*}
\centering
\epsscale{1.15}
\plotone{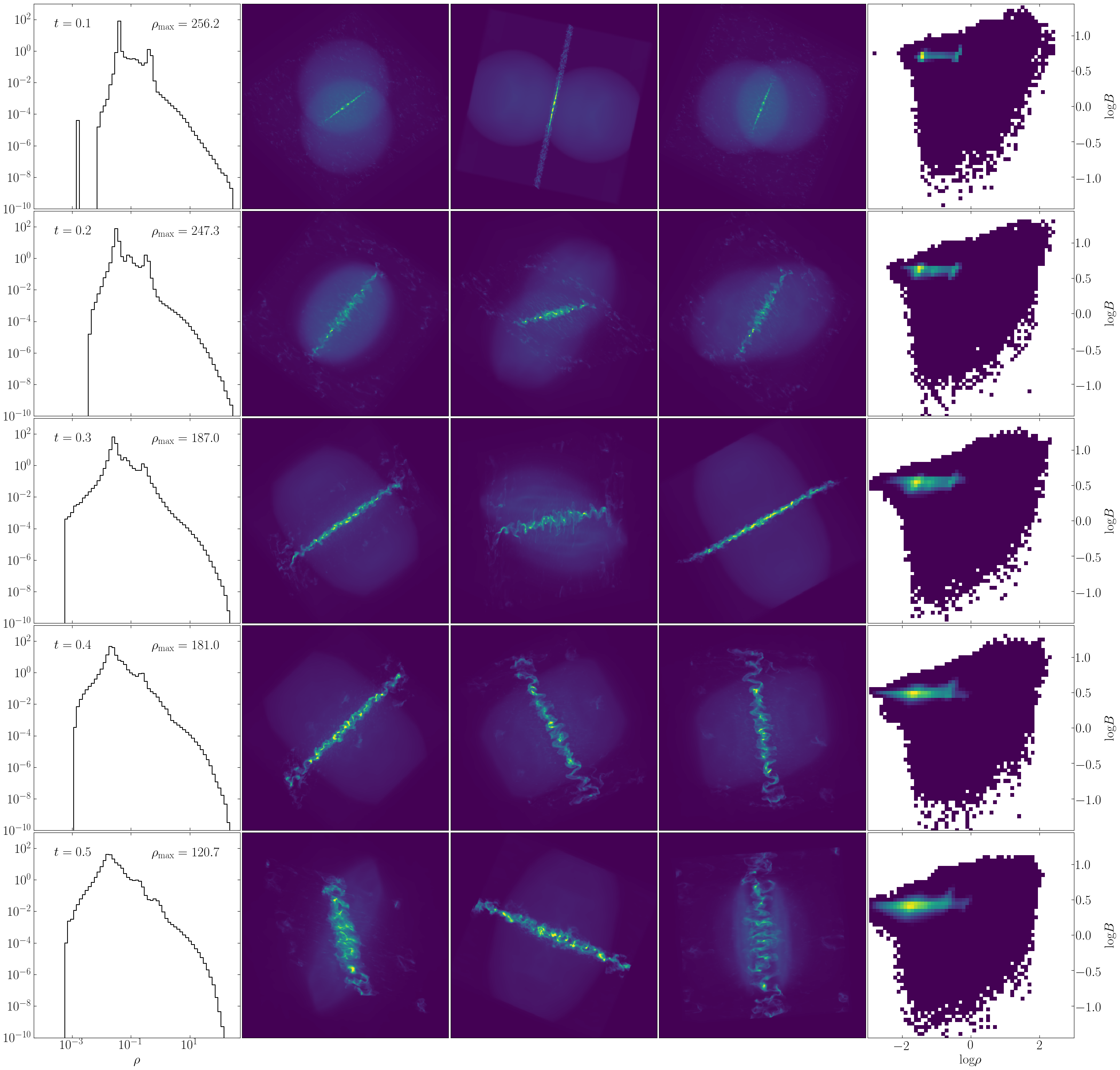}
\caption{
Same format as Figure \ref{fig:mrcol}, but for the \mbox{$B$\_H} model (\#4 in Table \ref{tab:ic}) and $B$ ranges from 0.03 to 30. 
\label{fig:b6}}
\end{figure*}

Figures \ref{fig:b1p5} and \ref{fig:b6} show the explorations for \mbox{$B$\_L} and \mbox{$B$\_H}, respectively. In general, \mbox{$B$\_L} has a lower maximum density (compared to \mbox{MRCOL}) throughout the simulation, except for t=0.3 when the maximum density reaches 173. At this time step, the high-density part of the $\rho$-PDF shows a linear shape which implies a power-law distribution. The $\rho$-PDF does not show much at low densities until t=0.4. At t=0.5, the maximum density of \mbox{$B$\_L} is about a factor of 2 smaller than that in \mbox{MRCOL}. Similar to \mbox{$\eta$\_H}, \mbox{$B$\_L} shows several large filaments around the main filament, and the overall morphology appears to be smoother than \mbox{MRCOL}. 

A noticeable feature is in the $B$-$\rho$ relation in which the bright pixels spread over a larger area. The initial two points preferentially populate the plane following a power-law relation $B\propto\rho^\alpha$, appearing as two diagonal lines in the plot. The gray dashed line in Figure \ref{fig:b1p5} shows a power-law with $\alpha=1$, which is approximately representative of the slope of the two diagonal lines. There is a third diagonal line between the two diagonal lines. With the progress of the simulation, the 3-diagonal feature becomes less prominent. At t=0.5, they smoothly extend to the vicinity, appearing as a bright blob in the $B$-$\rho$ plot. In fact, other explorations show the similar 3-diagonal feature, except that the feature is limited in a narrow $B$-range so not as prominent as in \mbox{$B$\_L}. The low initial B-magnitude allows the gas to dynamically evolve into a broader parameter space. The temporary 3-diagonal feature is probably a non-equilibrium phenomenon due to collision.

On the other hand, the \mbox{$B$\_H} exploration creates significantly denser gas. The maximum density reaches 256 at t=0.1 and continues to be $>$100 throughout the simulation. The $\rho$-PDF is also the broadest distribution so far, spanning a dynamic range of five orders of magnitude. The dense gas is easily visible in the RT images in which we see many bright clumps in the filament. The filament has rich sub-structures, including many curly fibers. We do not see the large parallel filaments around the main filament like those in \mbox{$B$\_L} and \mbox{$\eta$\_H}. The $B$-$\rho$ relation is not very different from \mbox{MRCOL}, except that the overall B-magnitude is higher by $\sim$2. The spread of B-magnitudes is narrower than that in \mbox{$B$\_L} in the bright pixels.

\subsection{Density}\label{subsec:dens}

\begin{figure*}
\centering
\epsscale{1.15}
\plotone{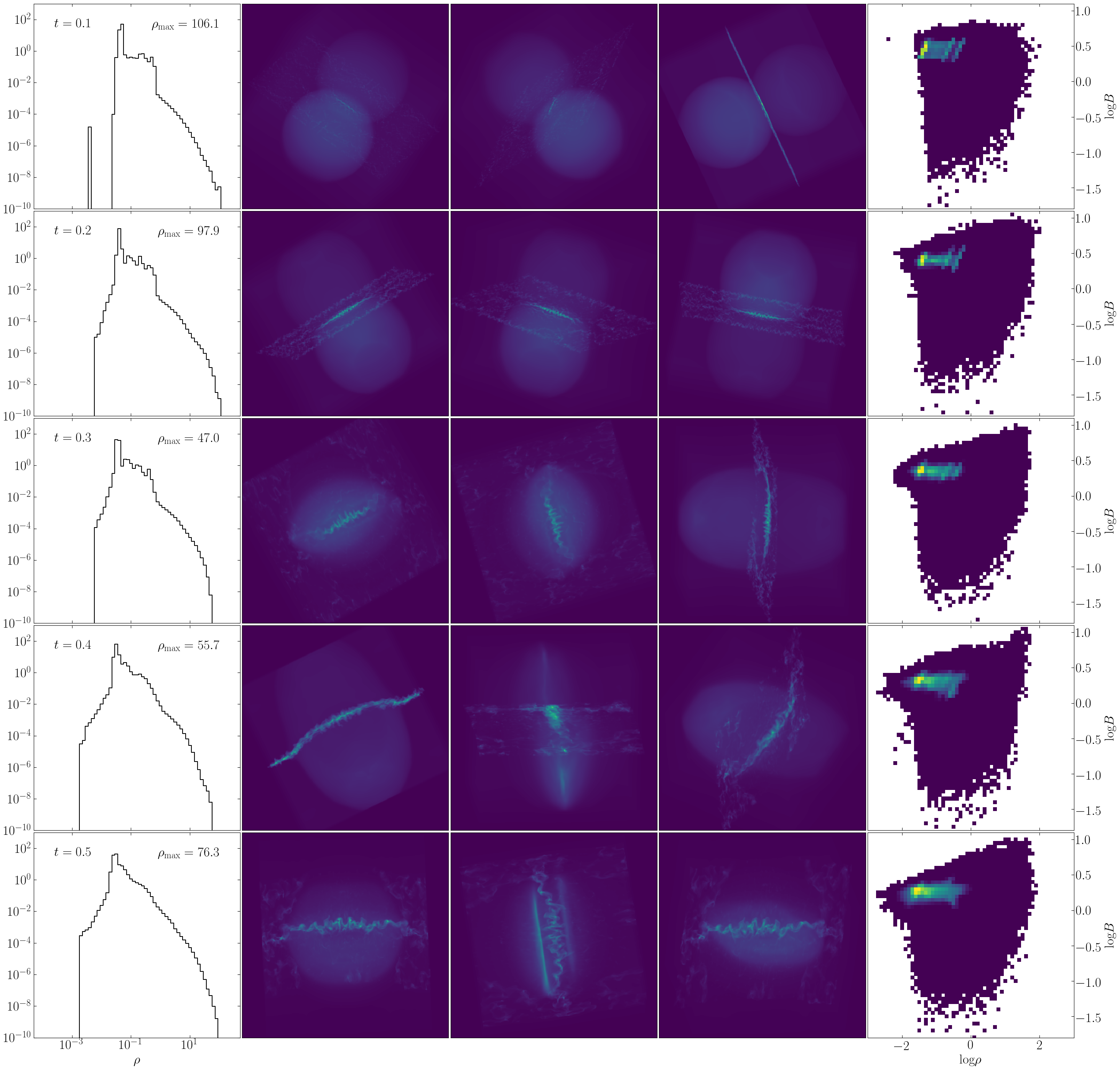}
\caption{
Same format as Figure \ref{fig:mrcol}, but for the \mbox{$\rho_2$\_L} model (\#5 in Table \ref{tab:ic}).
\label{fig:rhoB0p25}}
\end{figure*}

\begin{figure*}
\centering
\epsscale{1.15}
\plotone{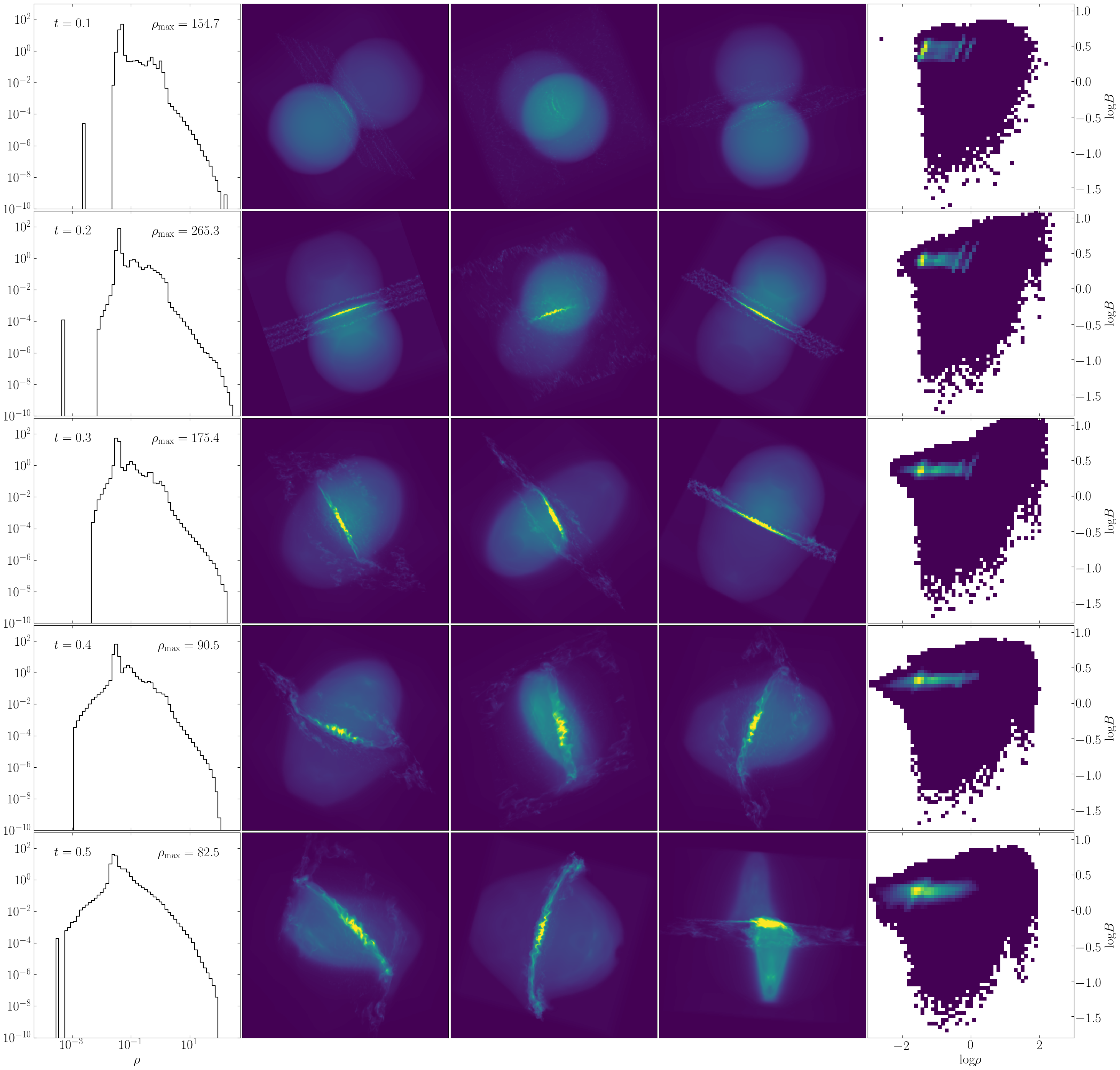}
\caption{
Same format as Figure \ref{fig:mrcol}, but for the \mbox{$\rho_2$\_H} model (\#6 in Table \ref{tab:ic}).
\label{fig:rhoB1p0}}
\end{figure*}

Figures \ref{fig:rhoB0p25} and \ref{fig:rhoB1p0} show the explorations \mbox{$\rho_2$\_L} and \mbox{$\rho_2$\_H}, respectively. Both simulations show that the filament becomes curved if the two colliding clouds have different densities. The curvature is simply because the denser cloud possesses a larger momentum. With the cloud collision and filament formation, the main filament protrudes into the cloud with lower density. These results clarify the misconception about K21 that CMR only produces ruler-straight filaments. 

As shown Figure \ref{fig:rhoB0p25}, the maximum density in the \mbox{$\rho_2$\_L} model is smaller than that in \mbox{MRCOL}, except at t=0.4. The filament morphology in \mbox{$\rho_2$\_L} is similar to that in \mbox{MRCOl} besides the curvature, i.e., with many curly fibers. The $B$-$\rho$ relation in \mbox{$\rho_2$\_L} is also not distinctive from that in \mbox{MRCOL}, except that the middle diagonal feature during early times seems closer to the higher density diagonal.

The \mbox{$\rho_2$\_H} model, however, shows some difference from the fiducial model. The maximum density in \mbox{$\rho_2$\_H} is higher than \mbox{MRCOL} except at t=0.5. Throughout the simulation, the $\rho$-PDF appears to be wider than that in \mbox{MRCOL}. The difference in $\rho$-PDF between the two models is likely due to the initial density diversity in \mbox{$\rho_2$\_H}. A notable difference of the morphology, besides the curvature, is that the main filament in \mbox{$\rho_2$\_H} shows a concentration of dense gas in the middle of the filament. In the $B$-$\rho$ plot, \mbox{$\rho_2$\_H} shows a low B-magnitude tail at high densities, which is not as obvious in other models. With VisIt, we confirm that the high-density low B-magnitude tail is in the middle part of the main filament. The $B$-$\rho$ relation at t=0.1 shows the 3-diagonal feature which is similar to the \mbox{\mbox{$B$\_L}} model (Figure \ref{fig:b1p5}).

\subsection{Cloud Radii}\label{subsec:radii}

\begin{figure*}
\centering
\epsscale{1.15}
\plotone{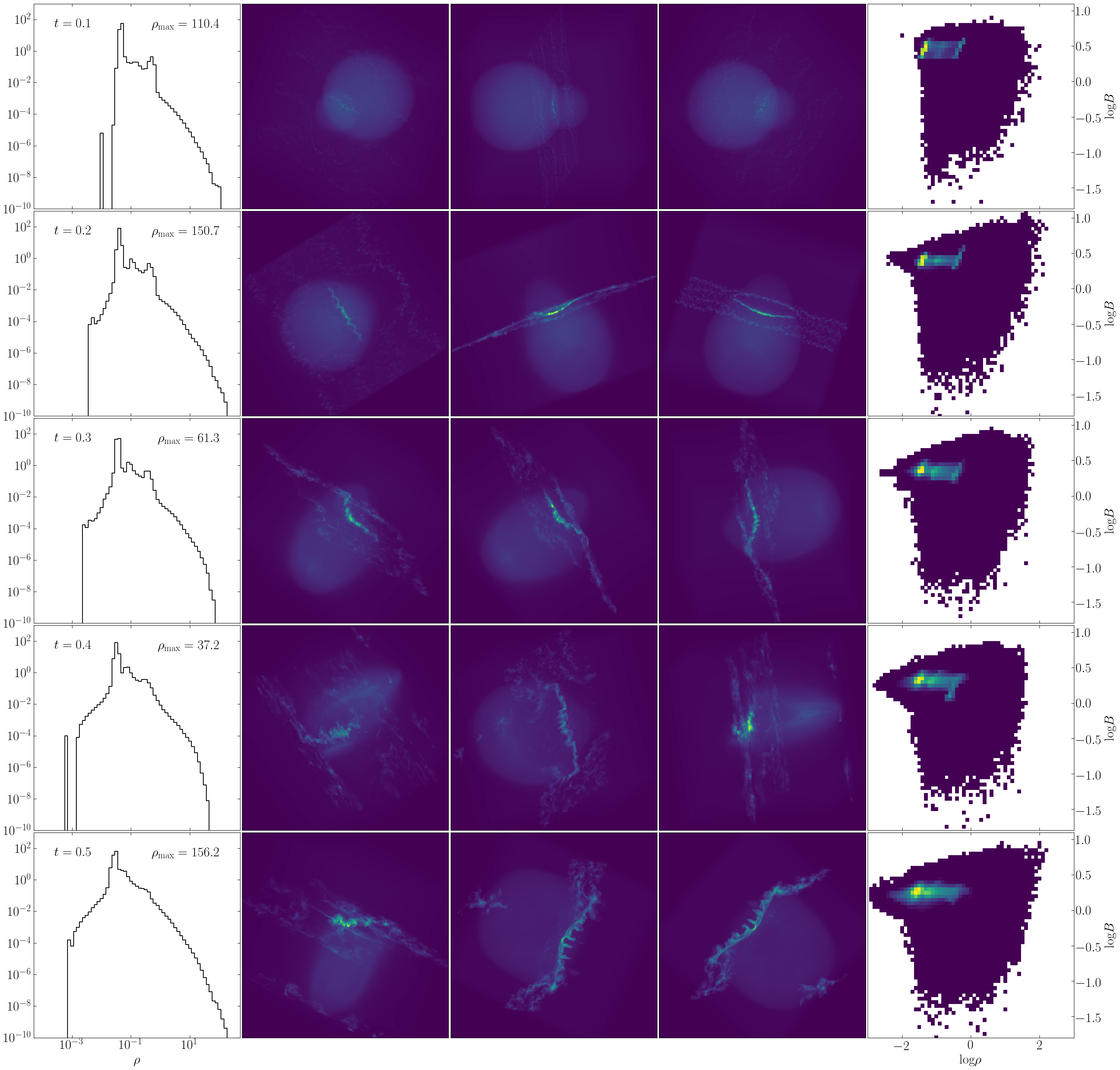}
\caption{
Same format as Figure \ref{fig:mrcol}, but for the \mbox{$R_2$\_L} model (\#7 in Table \ref{tab:ic}).
\label{fig:radAB0p45}}
\end{figure*}

\begin{figure*}
\centering
\epsscale{1.15}
\plotone{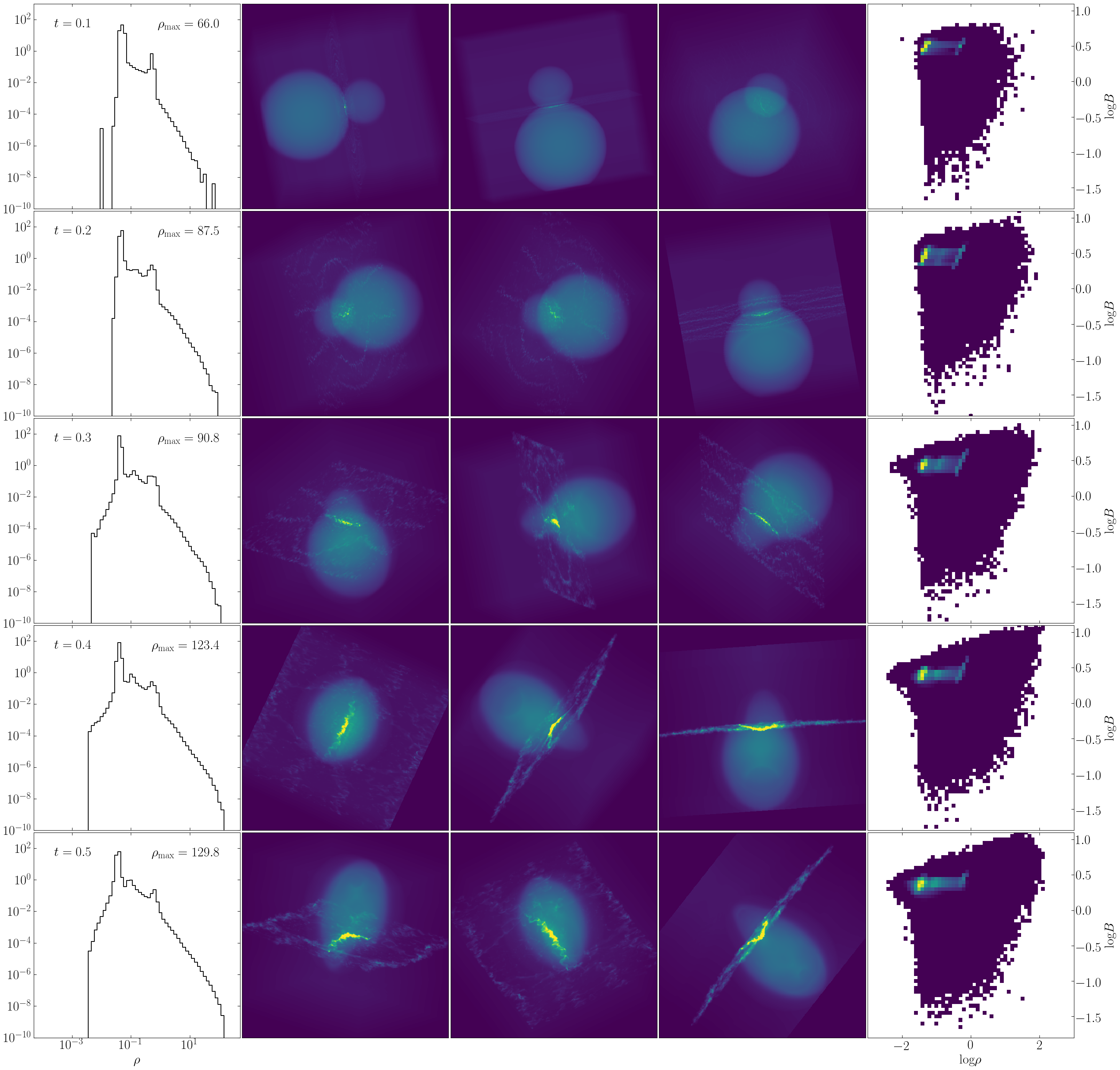}
\caption{
Same format as Figure \ref{fig:mrcol}, but for the \mbox{$R_2$\_H} model (\#8 in Table \ref{tab:ic}).
\label{fig:radAB1p8}}
\end{figure*}

Figures \ref{fig:radAB0p45} and \ref{fig:radAB1p8} show the explorations \mbox{$R_2$\_L} and \mbox{$R_2$\_H}, respectively. In the two simulations, we vary the second cloud radius by factors of 2 so that the colliding clouds have different radii. One can see that the differing sizes result in a curved main filament. The curvature forms because the smaller cloud plunges into the larger one. 

The main filament at t=0.5 in \mbox{$R_2$\_L} shows many spikes toward one side of the filament. In addition, we see accompanying large filaments around the main filament, similar to the results in \mbox{$B$\_L} and \mbox{$\eta$\_H}. The maximum density in \mbox{$R_2$\_L} fluctuates throughout the simulation. High-density cells concentrate in the middle of the main filament. The $B$-$\rho$ relation is not distinctive from that in the fiducial model \mbox{MRCOL}.

\mbox{$R_2$\_H}, however, does not show a strong fluctuation in the maximum density. Instead, it steadily increases from t=0.1 to t=0.5. Similar to \mbox{$R_2$\_L}, the high-density cells in \mbox{$R_2$\_H} also concentrate in the middle of the main filament. The $B$-$\rho$ plot shows some cells with very high B-magnitude ($>10$ or $>30\mu$G), although the overall $B$-$\rho$ relation shows no special features compared to \mbox{MRCOL}.

\subsection{Temperature}\label{subsec:T}

\begin{figure*}
\centering
\epsscale{1.15}
\plotone{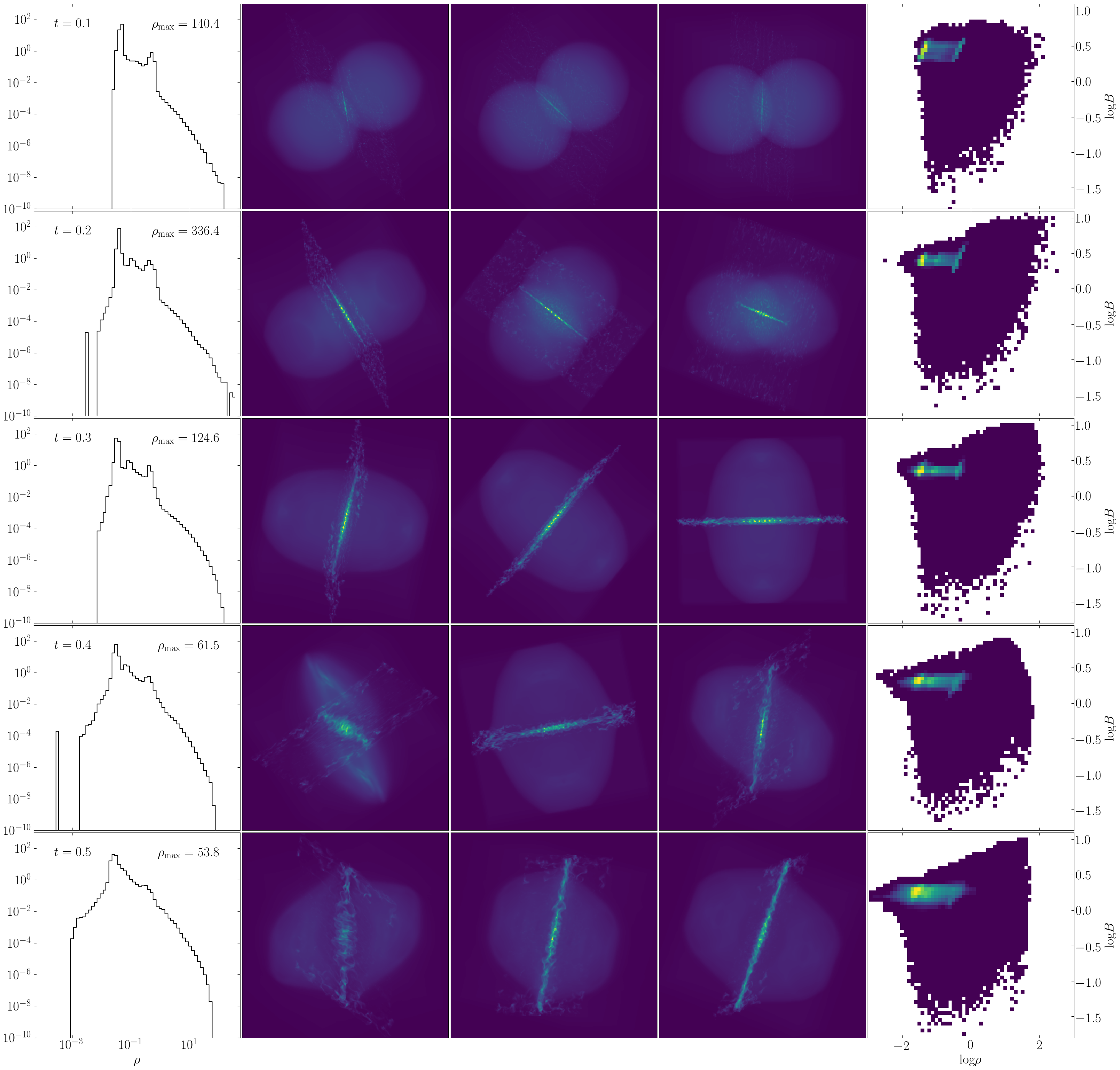}
\caption{
Same format as Figure \ref{fig:mrcol}, but for the \mbox{$T$\_L} model (\#9 in Table \ref{tab:ic}).
\label{fig:T10}}
\end{figure*}

\begin{figure*}
\centering
\epsscale{1.15}
\plotone{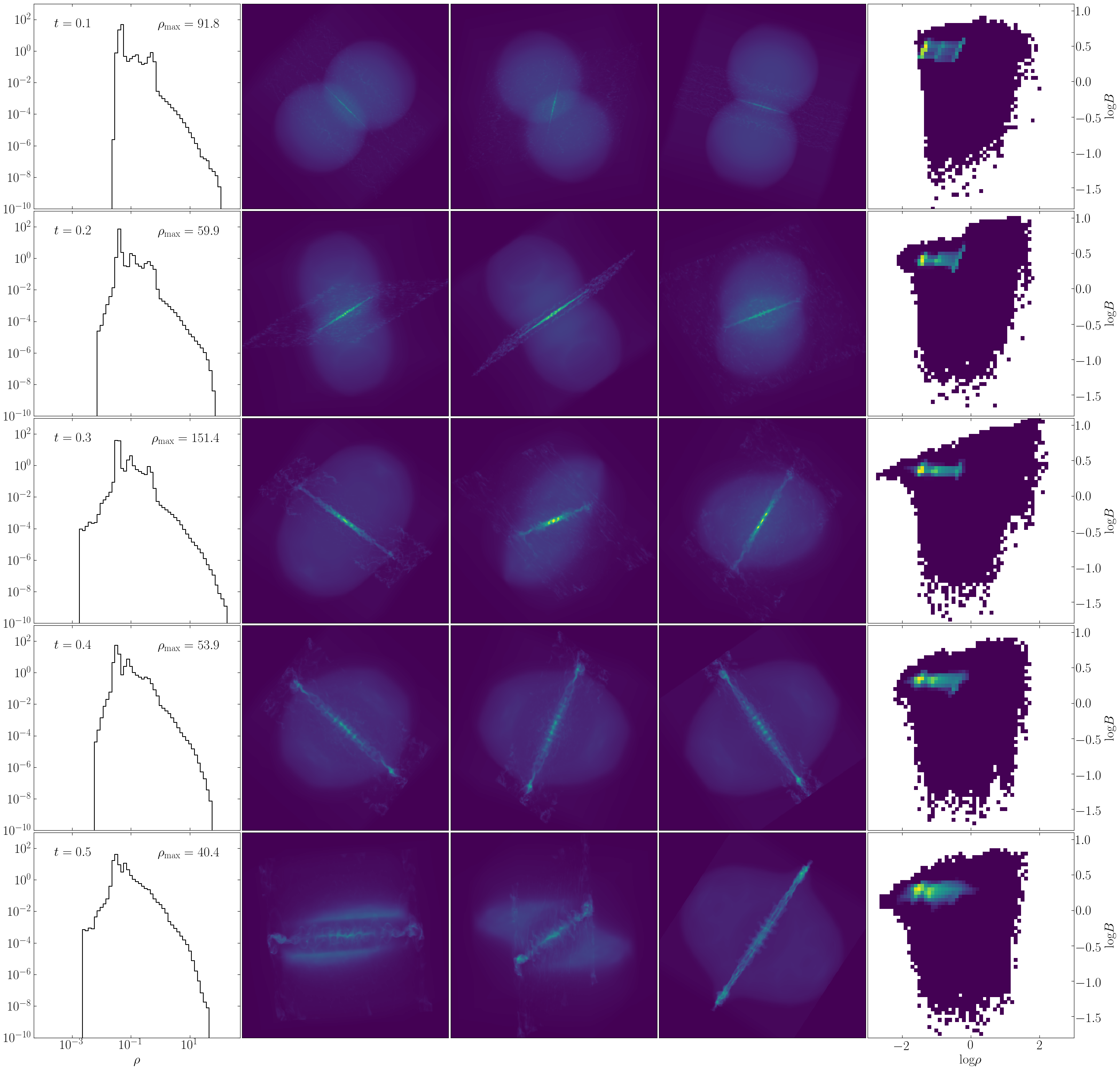}
\caption{
Same format as Figure \ref{fig:mrcol}, but for the \mbox{$T$\_H} model (\#10 in Table \ref{tab:ic}).
\label{fig:T30}}
\end{figure*}

Figures \ref{fig:T10} and \ref{fig:T30} show the explorations \mbox{$T$\_L} and \mbox{$T$\_H}, respectively. When the isothermal temperature is lower, the most noticeable difference is the creation of very high densities early in the simulation. As shown in Figure \ref{fig:T10}, the maximum density reaches $\sim$340 at t=0.2, which is a factor of $\sim$2.4 higher than that in the fiducial model \mbox{MRCOL}. A lower temperature means a lower plasma-$\beta$, i.e., the magnetic pressure becomes more important. A plausible scenario is that the lower thermal pressure allows the reconnected field to create denser gas in the filament. However, the field cannot hold the high densities long, as the maximum density drops to only $\sim$60 at t=0.4. The density drop implies that the filament mass is not large enough for self-gravity to dominate. So no gravitational collapse is triggered. However, the fiducial model, with a higher temperature, was able to create high densities again at t=0.5 after a brief drop at t=0.4 (Figure \ref{fig:mrcol}). At this point, we attribute this seemingly contradictory phenomenon to the highly stochastic nature of magnetic reconnection. More dedicated studies shall reveal the details in the future.

As shown in Figure \ref{fig:T30}, with a higher temperature in the \mbox{$T$\_H} model, CMR is generally less capable of producing high-density gas, although occasionally there is still a minor amount of such gas showing up in the $\rho$-PDF, e.g., at t=0.3. Again, this anomaly is probably due to the stochastic nature of magnetic reconnection. The morphology of the filament, as shown in the middle columns in Figure \ref{fig:T30}, shows no major difference from the fiducial model \mbox{MRCOL}. Neither does the $B$-$\rho$ plot. With even higher temperatures, we expect that it is harder for CMR to create dense gas. 

\subsection{Collision Velocity}\label{subsec:vcol}

\begin{figure*}
\centering
\epsscale{1.15}
\plotone{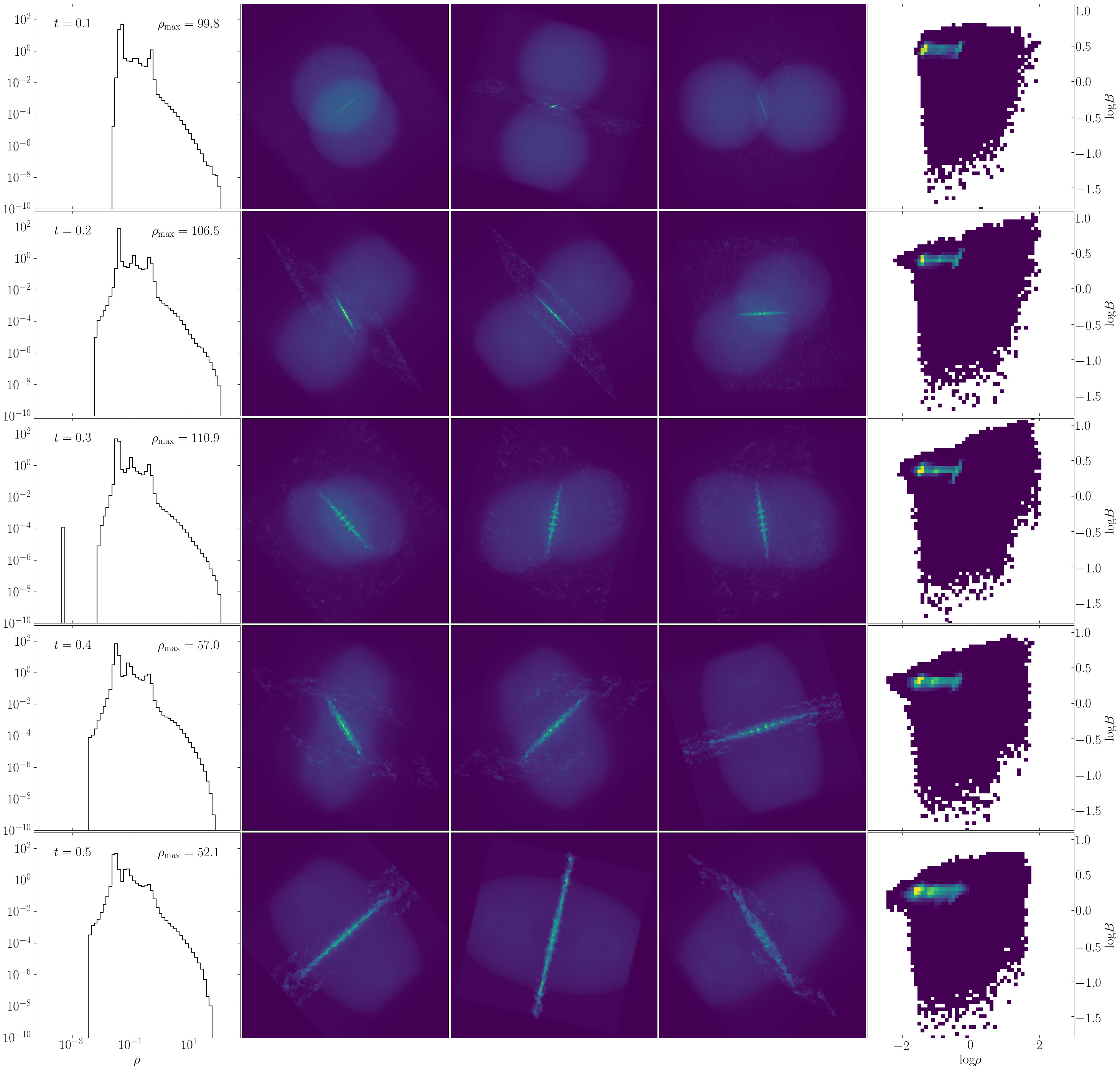}
\caption{
Same format as Figure \ref{fig:mrcol}, but for the \mbox{$v_x$\_L} model (\#11 in Table \ref{tab:ic}).
\label{fig:vcol1}}
\end{figure*}

\begin{figure*}
\centering
\epsscale{1.15}
\plotone{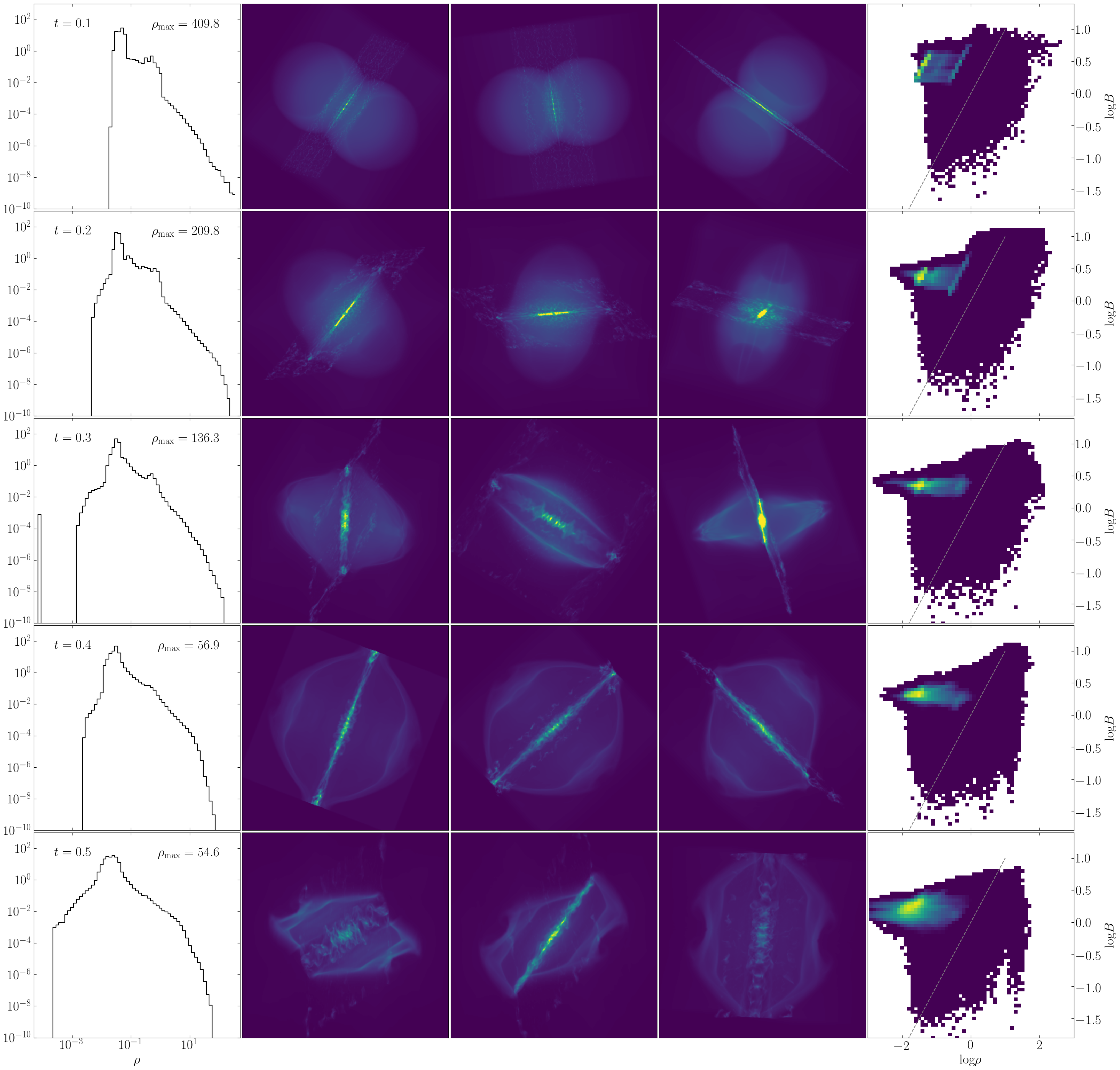}
\caption{
Same format as Figure \ref{fig:mrcol}, but for the \mbox{$v_x$\_H} model (\#12 in Table \ref{tab:ic}). The gray dashed line represents $B\propto\rho$ normalized at (0,0).
\label{fig:vcol4}}
\end{figure*}

Figure \ref{fig:vcol1} shows the exploration \mbox{$v_x$\_L}. With a slower collision (1 km s$^{-1}$), the gas density is generally smaller than \mbox{MRCOL}. The maximum density reaches 110 at t=0.3. The filament emission appears to be weaker than in \mbox{MRCOL} (Figure \ref{fig:mrcol}), although the filament morphology is not very different from the fiducial model. The $B$-$\rho$ plot in Figure \ref{fig:vcol1} shows no major difference compared to Figure \ref{fig:mrcol}, except for a lack of high-density and strong-field pixels.

Figure \ref{fig:vcol4} shows the exploration \mbox{$v_x$\_H}. The relative collision velocity is now 4 km s$^{-1}$ and the results show several noticeable features. The maximum gas density reaches 400 at t=0.1, which is a factor of 3 higher than \mbox{MRCOL}. The $\rho$-PDF shows a power-law tail at $\rho\ga1$. With the progress of the simulation, the maximum density drops from 200 at t=0.2 to 55 at t=0.5. One can see a turnaround in the $\rho$-PDF at $\rho\sim10$ at t=0.4 and t=0.5. Regarding the morphology, before t=0.3, we see bright emission from the high-density gas in the middle of the main filament. At t=0.5, the filament shows many curly fibers. The outer layer of the clouds show some density enhancement.

In the $B$-$\rho$ relation, we see that the high-speed collision creates a widespread bright pixels. At t=0.1, we see the 3-diagonal feature which is similar to that in the \mbox{$B$\_L} model (Figure \ref{fig:b1p5}). The feature persists at t=0.2, but dissolves at t$\ga$0.3. Again, we plot the $B\propto\rho$ relation as the gray line. One can see that the 3-diagonal feature is explicable with the linear relation $B\propto\rho$, just as that in the \mbox{$B$-L} model (Figure \ref{fig:b1p5}).

\subsection{Shear Velocity}\label{subsec:vshe}

\begin{figure*}
\centering
\epsscale{1.15}
\plotone{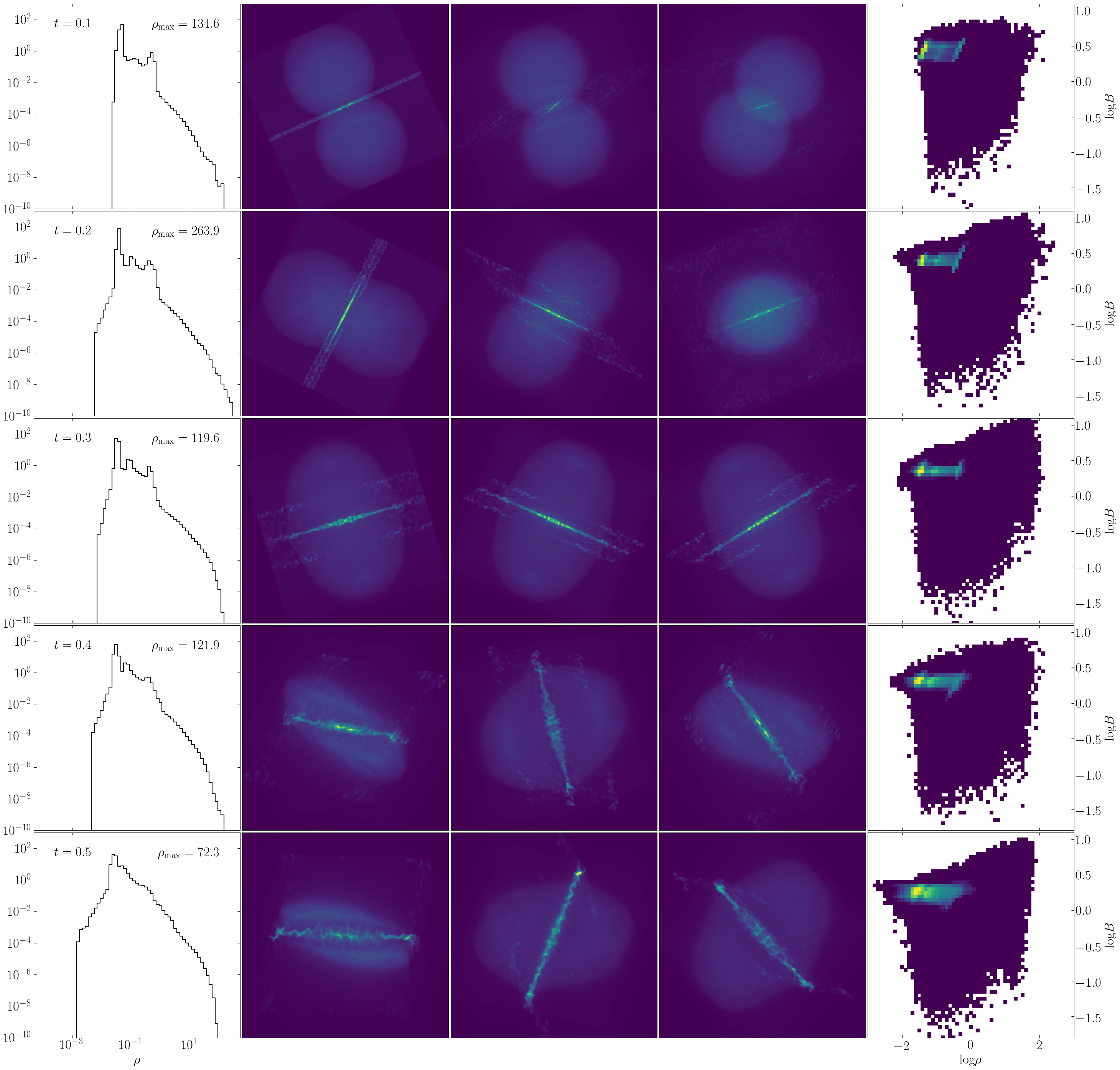}
\caption{
Same format as Figure \ref{fig:mrcol}, but for the \mbox{$v_z$\_L} model (\#13 in Table \ref{tab:ic}).
\label{fig:vshear0p25}}
\end{figure*}

\begin{figure*}
\centering
\epsscale{1.15}
\plotone{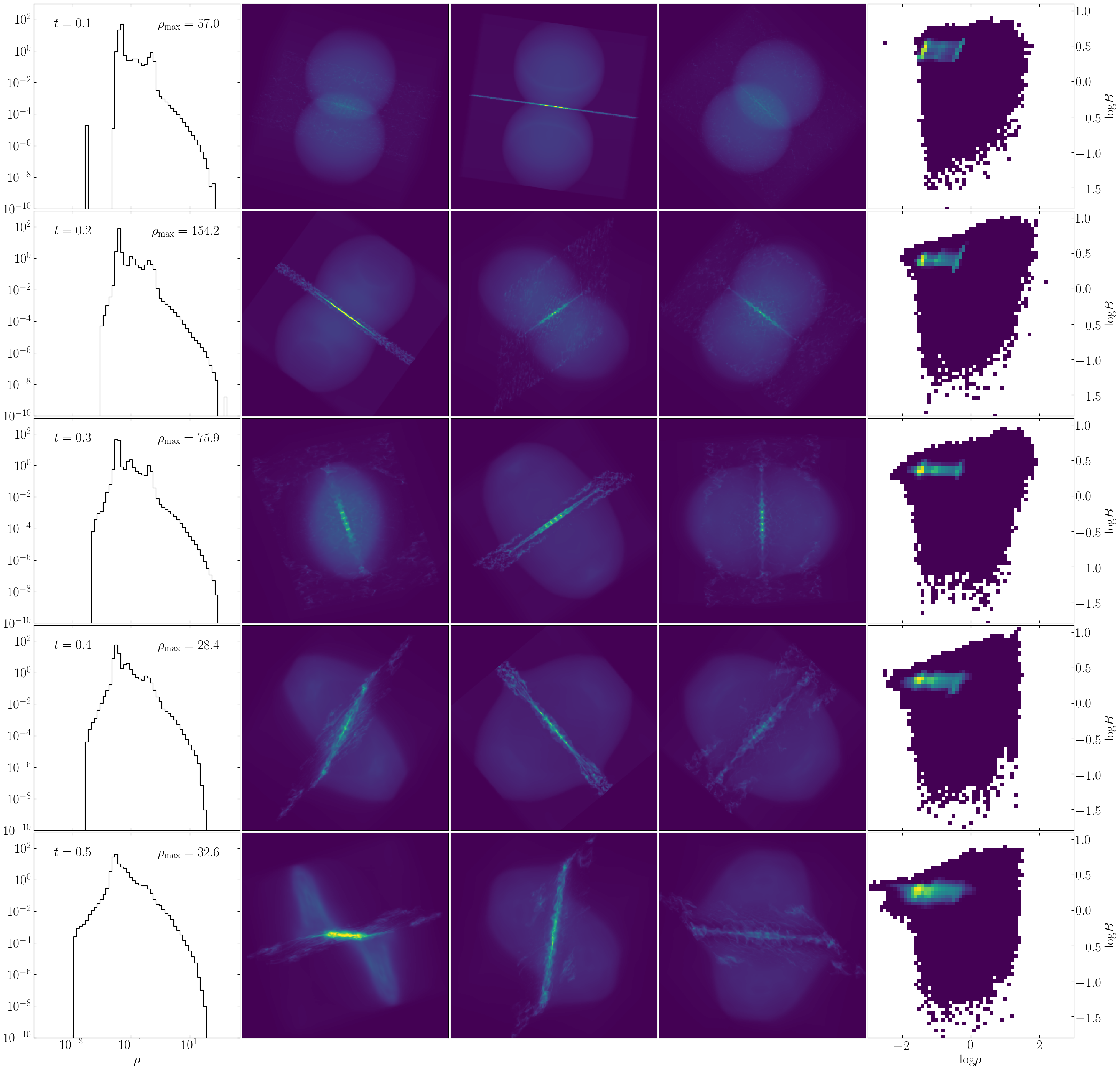}
\caption{
Same format as Figure \ref{fig:mrcol}, but for the \mbox{$v_z$\_H} model (\#14 in Table \ref{tab:ic}).
\label{fig:vshear1}}
\end{figure*}

Figure \ref{fig:vshear0p25} shows the exploration \mbox{$v_z$\_L} in which the shear velocity is reduced by a factor of 2 compared to \mbox{MRCOL}. The role of the shear velocity is to deposit an angular momentum into the filament. The faster the filament rotation, the less dense gas it holds, and vice versa. Therefore, it is understandable that the maximum density in \mbox{$v_z$\_L} is larger than \mbox{MRCOL} at t=0.2-0.4. Meanwhile, the morphology of the filament is similar to that in the fiducial model. The $B$-$\rho$ relation shows no major difference.

Figure \ref{fig:vshear1} shows the exploration \mbox{$v_z$\_H} in which the shear velocity is increased by a factor of 2. The increased shear velocity gives rise to more angular momentum in the filament. Therefore, we expect the filament to be less capable of holding dense gas. Indeed, the left column in Figure \ref{fig:vshear1} shows that the maximum density is significantly smaller than that in \mbox{MRCOL}, except for t=0.2 when there is only a minor amount of dense gas exceeding 100. The morphology of the filament shows no major difference from that in the fiducial model, except for an excess of curly fibers. The $B$-$\rho$ plot shows the 3-diagonal feature at t=0.1, which is less prominent at later times. At t=0.4-0.5, there is a lack of high-density pixels.

\subsection{Structure Characterization}\label{subsec:struc}

\begin{figure}[htb]
\centering
\includegraphics[angle=0,width=\columnwidth]{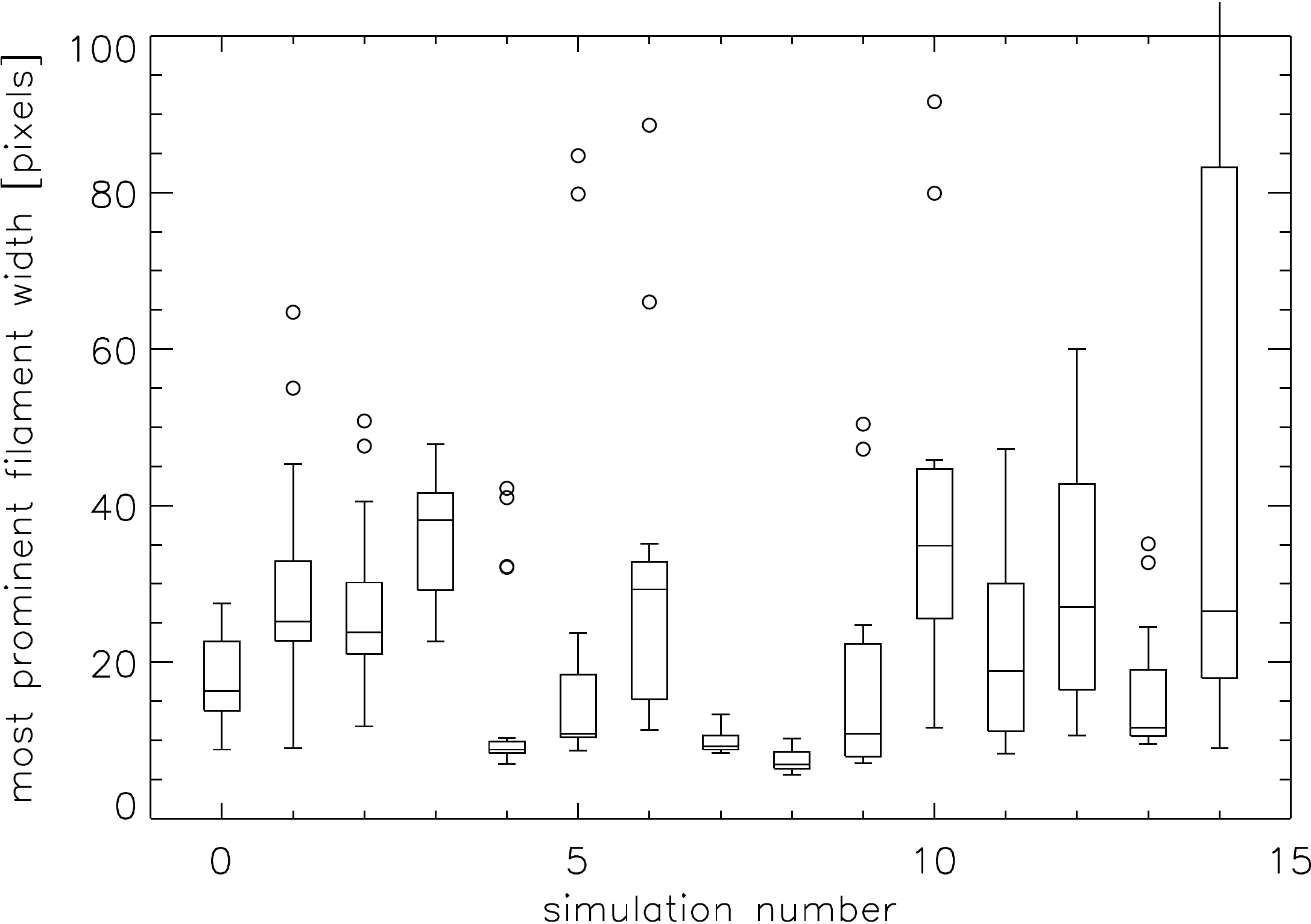}
\caption{Distribution of the most prominent structure widths detected in 20 random projections of the last timestep (1 Myr) of the different simulations through the anisotropic wavelet analysis. The distributions are given as whisker plots giving the two central quartiles of the distributions as boxes and the extremes (outer quartiles, like error bars) if they fall within 1.5 times the total span of the inner quartiles. Outliers beyond that limit are shown as individual circles. For simulation \#14 (\mbox{$v_z$\_H}) the upper quartile is truncated extending up to 165 pixels. 
\label{fig:widths}}
\end{figure}

Figure~\ref{fig:widths} shows the distribution of structure widths detected for the final time step ($t=0.5$) in all simulations when observing 20 different projections of that snapshot. The widths are measured through the anisotropic wavelet analysis \citep{OssenkopfOkadaStepanov2019} and are shown as whisker plots\footnote{http://www.idlcoyote.com/programs/cgboxplot.pro} showing the quartiles of the distributions and possible outliers. In most cases, the different projections are dominated by the main filament showing prominent widths in the range between about 15 and 30 pixels. Clear outliers are \#4 (\mbox{$B$\_H}, Figure \ref{fig:b6}), \#7 (\mbox{R$_2$\_L, Figure \ref{fig:radAB0p45}}), and \#8 (\mbox{R$_2$\_H}, Figure \ref{fig:radAB1p8}) with much narrower structures on the one hand and \#14 (\mbox{$v_z$\_H}, Figure \ref{fig:vshear1}) with many projections with a wider appearance on the other hand. In the three cases with narrow structures we can see by eye how the main filament breaks up into many substructures that are statistically dominating the projection although the main filament is still recognizable to some degree. In particular the \mbox{$B$\_H} simulation (\#4, Figure \ref{fig:b6}) shows some similarities to the stick filament with an overall straight structure composed of many inner clumps, rings, and forks. To a lesser degree this is also visible for the \mbox{$v_z$\_L} setup (\#13, Figure \ref{fig:vshear0p25}). In contrast the \mbox{$v_z$\_H} simulation (\#14, Figure \ref{fig:vshear1}) shows not only one filament, but three prominent elongated structures that merge into one prominent structure width under several projection angles giving this very wide distribution of widths that is basically unusable to characterize the underlying structure in terms of filament properties. To some degree this superposition also occurs for \#1 (\mbox{$\eta$\_L}, Figure \ref{fig:re0p0001}), \#2 (\mbox{$\eta$\_H}, Figure \ref{fig:re0p01}), and \#12 (\mbox{$v_x$\_H}, Figure \ref{fig:vcol4}) explaining outliers and high values there.

\begin{figure}[htb]
\centering
\includegraphics[angle=0,width=\columnwidth]{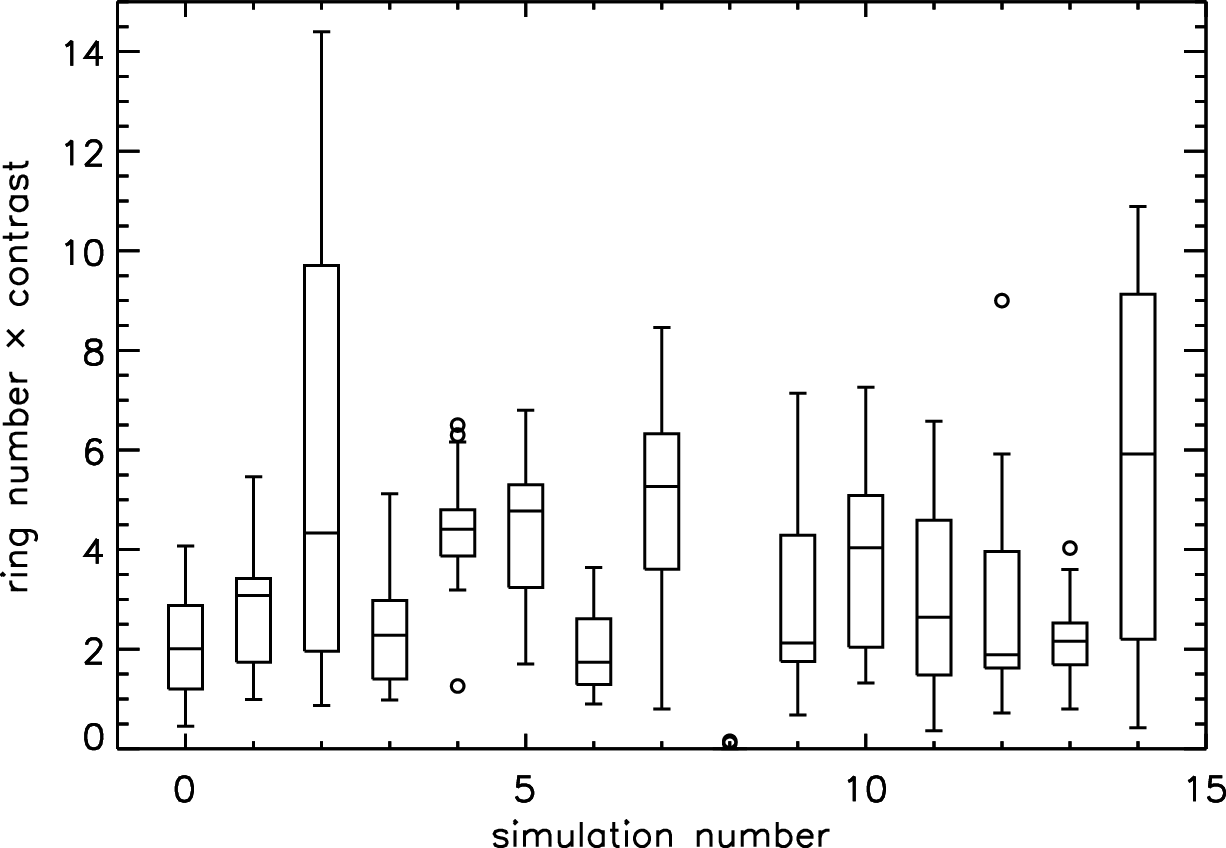}
\caption{Distribution of the ``ringiness'' parameter for the 20 random projections of the last timestep of the different simulations. The parameter gives the number of detected rings with inner radii between 5 and 20 pixels and a radius ratio between 1.2 and 1.4 multiplied with their contrast relative to a filled circle with the same outer radius. The distributions are given as whisker plots like in Figure~\ref{fig:widths}. 
\label{fig:allrings}}
\end{figure}

Figure~\ref{fig:allrings} shows the distribution of the measured number and contrast of ring and fork structures detected in the filaments for the final time step ($t=0.5$). The projected images were scanned for rings with inner radii between 0.035 and 0.1~pc and ratios between outer and inner radius between 1.2 and 1.4. 
There is only one simulation, \mbox{$R_2$\_H}, where no rings are detected in any projection. For most other simulations we find a typical ``ringiness'' parameter of 2-3. Significant deviations from this behavior are found for the simulations \#2, \#7, and \#14 that have larger medium ringiness parameters but also a significantly larger scatter between the different projections. Some projections show almost no rings while others show a very fuzzy filament structure. This is very prominent for the very flat filament in the \mbox{$\eta$\_L} simulation in Figure~\ref{fig:re0p0001}. This is actually an effect that we expect from CMR because the B-field is not isotropic but prefers one plane for the ring formation. A different behavior is found for the simulations \#4 and \#5 that also show more rings than most other simulations, but no increased spread between different projections. This is confirmed when inspecting Figures \ref{fig:b6} and \ref{fig:rhoB0p25}. They show that the filament is broken up into many substructures in all projections describing the overall shape rather as a narrow spiral than a filament. Consequently we can use Figure~\ref{fig:allrings} to categorize the way the CMR filament breaks up into substructure depending on the parameters of the clouds and their collision.

\section{Discussion}\label{sec:discus}

The parameter exploration in \S\ref{sec:results} give us an initial census of the CMR behavior in response to different conditions. In general, CMR can produce denser gas if the initial B-field is stronger, or the initial cloud density is higher, or the temperature is lower, or the collision velocity is higher, or the shear velocity is lower. The latter four conditions are not unexpected. For example, a higher temperature implies a higher thermal pressure and a higher shear velocity gives rise to a larger angular momentum. The higher pressure and angular momentum give more support to the filament and thus suppress dense gas formation. 

Traditionally, B-fields are thought to be a factor that supports molecular gas against collapse. So intuitively one expects less dense gas in the presence of strong B-fields. In the context of CMR, the seemingly contradictory phenomenon actually makes perfect sense, because B-fields are no longer a passive resisting factor but an active force that produces dense gas. As shown in \S\ref{sec:intro} and Figure \ref{fig:stream}, the dense filament forms due to the confining force from the reconnected fields. So the stronger the B-field, the more capable it is to confine more gas into the central filament, thus producing more denser gas. In this sense, an initially magnetically sub-critical volume of gas is more likely to form a dense filament if it is hit by another volume of gas that carries reverse fields. 

Our parameter study also provides guidance for distinguishing CMR-filaments in observations. First of all, a necessary condition for CMR is the reverse B-field. Once the filament forms, it is surrounded by a helical field (Figure \ref{fig:stream}). Therefore, if we see reverse fields around a molecular cloud in observation, the cloud can be a candidate of a CMR-filament. Such a cloud should also have a perpendicular plane-of-sky B-fields as can be deduced via dust polarization mapping. A nice example is the famous Orion A cloud. Over a course of 15-year-long observation, \citet{1997ApJS..111..245H} mapped the line-of-sight B-field component in a vast $20^\circ\times20^\circ$ area across the Orion cloud complex, corresponding to a 140 pc $\times$ 140 pc region at the distance of 400 pc. The extraordinary work showed that the Orion A cloud is right in between a reverse B-field. Note, this field-reversal is over a large area. It is not a small-scale fluctuation. Later, the Planck polarization observation showed a perpendicular plane-of-sky B-field relative to the Orion A main filament. Therefore, the Orion A cloud is an excellent CMR-filament candidate. Recently, through Faraday Rotation measurements, \citet{2020IAUGA..30..103T,2022A&A...660A..97T} found that the Perseus cloud and the California cloud are also in between reverse B-fields. This finding indicates that CMR may be more prevalent than we think. 

The summary of the CMR-filament properties (Figures \ref{fig:widths} and \ref{fig:allrings}) can further help us identify the candidates. For instance, model \#4 (\mbox{$B$\_H}, Figure \ref{fig:b6}) preferentially has thin fibers (also model \#7 \mbox{$R_2$\_L} and model \#8 \mbox{$R_2$\_H}). Therefore, when we see a CMR-filament candidate with rich substructures that are narrow fibers, we know the cloud is probably strongly magnetized in the context of CMR. Interestingly, with high-resolution observations, \citet{2018A&A...610A..77H} showed widespread fibers in the Orion A filament, which is consistent with the earlier speculation that Orion A was a result of CMR. Before that, \citet{2017A&A...606A.123H} also observed fibers in the Perseus cloud (NGC 1333). On the other hand, a lower initial B-magnitude (model \#3 \mbox{$B$\_L}, Figure \ref{fig:b1p5}) typically results in larger widths (Figure \ref{fig:widths}). Similar results are also achievable with higher temperature (\#10). 

The rich substructures in a CMR-filament appear as rings and forks in RT images (K21). Therefore, the evaluation of the ringiness (\S\ref{subsec:struc}) provides a useful characterization of the structural organization in CMR-filaments, which is ultimately determined by the underlying physics. Figure \ref{fig:allrings} shows that smaller cloud sizes (\#7) and higher shear velocity (\#14) appear to be more ringy. If the medium is highly resistive (\#2), the ring-morphology becomes more diverse (also true for \#7). On the contrary, larger cloud sizes (\#8) suppress the formation of rings, which is also the case with high initial densities (\#6) and low shear velocity (\#13). Interestingly, both model \#6 and model \#8 form a curved CMR-filament. The fact that both models suppress rings indicates that the ring formation needs the axisymmetry. Other models that are more ringy include \#4 (\mbox{$B$\_H}) and \#5 (\mbox{$\rho_2$\_L}).

Note that the above discussion about width and ringiness was based on a single time step (t=0.5). As mentioned earlier, the CMR process is highly variable both in space and time. So the two characteristics could vary significantly. In fact, the actually collision could be more complicated than two spherical clouds. One can imagine two large planes running into each other from afar, potentially due to supernova explosion, massive star wind, or gas flow from high Galactic latitude (even outside the Galaxy). If the two planes have a bumpy surface with (perturbed) B-fields, their collision will create a web of CMR-filaments (or irregular dense gas if the fields are not exactly antiparallel, see K21). All CMR needs are the reverse B-field, the protruding geometry, and the collision. Depending on the initial cloud arrangement and field orientation, the resulting dense gas at the field-reversal interface will vary significantly. More numerical studies will clarify.

\section{Summary}\label{sec:summary}

In this paper, we have carried out a pilot exploration of CMR as a function of initial conditions. In particular, we vary the Ohmic resistivity (models \mbox{$\eta$\_L} and \mbox{$\eta$\_H}), the initial magnetic field (models \mbox{$B$\_L} and \mbox{$B$\_H}), the initial cloud density (models \mbox{$\rho_2$\_L} and \mbox{$\rho_2$\_H}), the initial cloud radius (models \mbox{$R_2$\_L} and \mbox{$R_2$\_H}), the isothermal temperature (models \mbox{$T$\_L} and \mbox{$T$\_H}), the collision velocity (models \mbox{$v_x$\_L} and \mbox{$v_x$\_H}), and the shear velocity (models \mbox{$v_z$\_L} and \mbox{$v_z$\_H}) to investigate how they change the resulting CMR-filament properties, including the $\rho$-PDF, the filament morphology, and the $B$-$\rho$ relation. For this pilot exploration, we focus on the morphology in the far infrared dust emission at 250 $\mu$m wavelength. We have shown the 3D distribution of density and magnetic fields. We have summarized the time-dependent filament properties for each parameter exploration.

Compared to the fiducial model \mbox{MRCOL}, a stronger magnetic field (\mbox{$B$\_H}), or a higher initial density (\mbox{$\rho_2$\_H}), or a lower temperature (\mbox{$T$\_L}), or a higher collision velocity (\mbox{$v_x$\_H}), or a lower shear velocity (\mbox{$v_z$\_L}) produces denser gas. The strong field condition is seemingly counterintuitive, but it manifests the nature of the CMR process in which magnetic fields proactively creates dense gas instead of passively hindering its formation. At different time steps, other models can generate denser gas than some of the above five models. In fact, several models show a fluctuation at high densities, which is shown by the fluctuating $\rho$-PDF tail through the five output time steps. The density fluctuation shows the chaotic nature of magnetic reconnection, which is clearly seen in the filament morphology. Specifically, the CMR-filaments show plenty of irregular sub-structures, including curly fibers, rings, forks, and spikes. They come and go with time, and constitute complex density structures that vary with projection. The complex field topology around the filament (Figure \ref{fig:stream}) likely causes the rich structures.

With the progress of simulation, the $B$-$\rho$ plane is quickly (within 0.2 Myr) populated from just two points initially (the cloud density and the ambient density with the same B-field strength). During early times, most models show two parallel diagonal features extending from the two points, with a possible third diagonal feature in between. In logarithmic scale, the 3-diagonal feature indicates a power-law relation, which can be approximately represented by $B\propto\rho$. The diagonal extension varies by models, and only holds up to $\sim$0.6 Myr until it dissolves into a smooth distribution, which appears to be horizontal (constant $B$) in the $B$-$\rho$ plane. The temporary diagonal feature probably shows a non-equilibrium process during CMR.

To statistically characterize the CMR-filament structure, we develop two methods to quantify the distribution of fibers and rings within the filament. We perform a wavelet decomposition of the synthetic observation at different projections to assess the dominating fiber width. We then use a similar approach but with additional selection criteria to compute a ringiness parameter. We find that a stronger magnetic field (\mbox{$B$\_H}) or a differing cloud radius (\mbox{$R_2$\_L} and \mbox{$R_2$\_H}) tends to have thinner fibers, while a larger initial cloud scale (\mbox{$R_2$\_H}) produces negligible rings. 

\acknowledgments 
We thank an anonymous referee for the constructive report. We thank the Yale Center for Research Computing for guidance and use of the research computing infrastructure, specifically the Grace cluster. V.O. acknowledges support through the Collaborative Research
Centre 956, subproject C1, funded by the Deutsche Forschungsgemeinschaft (DFG), project ID 184018867, and support for the Article Processing Charge from the DFG (German Research Foundation, 491454339). 
H.G.A. acknowledges support from the National Science Foundation award AST-1714710. 
R.S.K. acknowledges financial support from the European Research Council via the ERC Synergy Grant ``ECOGAL'' (project ID 855130), from the Deutsche Forschungsgemeinschaft (DFG) via the Collaborative Research Center ``The Milky Way System''  (SFB 881 -- funding ID 138713538 -- subprojects A1, B1, B2 and B8), from the Heidelberg Cluster of Excellence (EXC 2181 - 390900948) ``STRUCTURES'', funded by the German Excellence Strategy, and from the German Ministry for Economic Affairs and Climate Action in project ``MAINN'' (funding ID 50OO2206). R.S.K. also thanks for computing resources provided by the Ministry of Science, Research and the Arts (MWK) of the State of Baden-W\"{u}rttemberg through bwHPC and DFG through grant INST 35/1134-1 FUGG and for data storage at SDS@hd through grant INST 35/1314-1 FUGG.
D.X. acknowledges support from the Virginia Initiative on Cosmic Origins (VICO). 

\software{Matplotlib \citep{matplotlib}, Numpy \citep{numpy}, SAOImageDS9 \citep{2003ASPC..295..489J}, VisIt \citep{HPV:VisIt}}

\bibliography{ref}
\bibliographystyle{aasjournal}

\end{document}